\documentclass[12pt]{iopart}

\usepackage{cite}
\usepackage{enumerate}
\usepackage{amssymb}
\usepackage{amsbsy}
\usepackage[dvips]{graphicx}
\usepackage{color}
\usepackage{graphics}

\newcommand{\be}{\begin{equation}} \newcommand{\ee}{\end{equation}}
\newcommand{\bea}{\begin{eqnarray}} \newcommand{\eea}{\end{eqnarray}}
\newcommand{\el}{\nonumber \\}
\newcommand{\re}[1]{(\ref{#1})}

\newcommand{\pat}{\partial}
\newcommand{\abs}[1]{|#1|}
\renewcommand{\sec}[1]{section \ref{#1}}
\newcommand{\fig}[1]{figure \ref{#1}}
\newcommand{\brt}[1]{[#1]}
\newcommand{\para}{\paragraph}

\renewcommand{\a}{\alpha}
\renewcommand{\b}{\beta}
\renewcommand{\c}{\gamma}
\renewcommand{\d}{\delta}
\newcommand{\mpc}{\mbox{$h^{-1}$Mpc}}
\newcommand{\LCDM}{$\Lambda$CDM\ }
\newcommand{\GN}{G_{\mathrm{N}}}
\newcommand{\ha}{\frac{1}{2}}
\newcommand{\erf}{\mathrm{erf}}
\newcommand{\keq}{k_{\mathrm{eq}}}

\newcommand{\teq}{t_{\mathrm{eq}}}

\newcommand{\afrw}{a_{\mathrm{EdS}}}
\newcommand{\Hfrw}{H_{\mathrm{EdS}}}
\newcommand{\fb}{f_{\mathrm{b}}}

\newcommand{\adot}{\dot{a}}
\newcommand{\addot}{\ddot{a}}
\newcommand{\rhodot}{\dot{\rho}}

\renewcommand{\H}{\frac{\adot}{a}}
\newcommand{\HH}{\frac{\adot^2}{a^2}}

\newcommand{\phit}{\tilde{\phi}}

\newcommand{\av}[1]{\langle{#1}\rangle}
\newcommand{\sQ}{\mathcal{Q}}
\newcommand{\sR}{{^{(3)}R}}
\newcommand{\avR}{\av{\sR}}
\newcommand{\Om}{\Omega_{\mathrm{m}}}
\newcommand{\Omn}{\Omega_{\mathrm{m0}}}
\newcommand{\om}{\omega_{\mathrm{m}}}
\newcommand{\Ob}{\Omega_{\mathrm{b}}}
\newcommand{\OQ}{\Omega_{\sQ}}

\newcommand{\OR}{\Omega_{R}}

\newcommand{\PRD}[1]{{\it Phys. Rev.} {\bf D#1}}
\newcommand{\PRE}[1]{{\it Phys. Rev.} {\bf E#1}}
\renewcommand{\PRL}[1]{{\it Phys. Rev. Lett.} {\bf #1}}

\newcommand{\NPB}[1]{{\it Nucl. Phys.} {\bf B#1}}
\newcommand{\PLA}[1]{{\it Phys. Lett.} {\bf A#1}}
\newcommand{\PLB}[1]{{\it Phys. Lett.} {\bf B#1}}
\newcommand{\MNRAS}[1]{{\it Mon. Not. Roy. Astron. Soc.} {\bf #1}}
\newcommand{\APJ}[1]{{\it Astrophys. J.} {\bf #1}}
\newcommand{\APJS}[1]{{\it Astrophys. J. Suppl.} {\bf #1}}

\renewcommand{\CQG}[1]{{\it Class. Quant. Grav.} {\bf #1}}
\newcommand{\GRG}[1]{{\it Gen. Rel. Grav.} {\bf #1}}
\newcommand{\AaA}[1]{{\it Astron. \& Astrophys.} {\bf #1}}
\newcommand{\PROG}[1]{{\it Prog. Theor. Phys.} {\bf #1}}
\newcommand{\AJ}[1]{{\it Astron. J.} {\bf #1}}
\newcommand{\IJMPD}[1]{{\it Int. J. Mod. Phys.} {\bf D#1}}

\begin{document}

\begin{titlepage}

\title{Evaluating backreaction with the peak model of structure formation}

\author{Syksy R\"{a}s\"{a}nen}

\address{Universit\'e de Gen\`eve, D\'epartement de Physique Th\'eorique\\
24 quai Ernest-Ansermet, CH-1211 Gen\`eve 4, Switzerland}

\ead{syksy {\it dot} rasanen {\it at} iki {\it dot} fi}

\begin{abstract}

\noindent
We evaluate the average expansion rate of a universe
which contains a realistic evolving ensemble of non-linear
structures.
We use the peak model of structure formation to
obtain the number density of structures, and take
the individual structures to be spherical.
The expansion rate increases relative to the FRW
value on a timescale of 10--100 billion years,
because the universe becomes dominated by fast-expanding voids.
However, the increase is not rapid enough to correspond
to acceleration.
We discuss how to improve our treatment.
We also consider various qualitative issues related
to backreaction.

\end{abstract}


\end{titlepage}

\setcounter{tocdepth}{2}

\tableofcontents

\setcounter{secnumdepth}{3}

\section{Introduction} \label{sec:intro}

\paragraph{Observations and acceleration.}

There is overwhelming observational evidence against homogeneous
and isotropic models of the universe based on standard general
relativity (i.e. the Einstein-Hilbert action in four dimensions)
with ordinary matter (i.e. baryons, dark matter, neutrinos and photons).
Observations are usually interpreted keeping to the linearly
perturbed homogeneous and isotropic Friedmann-Robertson-Walker
(FRW) models, in which case it is necessary to modify gravity
or introduce matter with negative pressure.
The \LCDM model, which involves the simplest modification
of gravity, the cosmological constant, or the equivalent form of
matter, vacuum energy, is in agreement with various observations of
cosmological distances. In FRW models, distance has a simple
correspondence with the expansion rate (and spatial curvature).

However, linearly perturbed FRW models do not describe
the non-linear structures present in the real universe.
It is also important to recognise that the only
evidence for modified gravity or exotic matter comes from
observations of cosmological distances, interpreted in terms of the
expansion rate. For example, deviations from general relativity
have not been observed in the solar system \cite{Reynaud:2008}.
(The Pioneer anomaly is a possible exception. However, it does
not agree with the predictions of the modified gravity models
which have been developed to explain the cosmological observations.)

The situation is quite different from that of dark matter,
for which there are several independent lines of evidence, such
as the motions of stars in galaxies, the motions of clusters,
the peak structure of the cosmic microwave background (CMB),
the early formation of structures, gravitational lensing,
as well as direct measures of the matter density
combined with the baryon density given by Big Bang Nucleosynthesis.
For this reason, constructing alternatives to dark matter requires
resort to baroque models \cite{teves}, if it is possible at all.
In contrast, in order to explain the observations without
modified gravity or exotic matter, it is only necessary
to change the distance scale (or the expansion rate)
as a function of redshift.

The measured quantities from which the expansion rate is inferred
can be divided into background quantities and perturbations,
though measurements of background quantities also involve
perturbations, apart from the luminosities of Type Ia supernovae (SNe Ia).
(In realistic models with non-linear structures, there is no simple
division into background and perturbations.)
Most of the information on the expansion rate comes from
measures of the background geometry. While the expansion rate
has been measured from different physical systems, such as
the CMB \cite{Spergel:2006}, large scale structure \cite{BAO}
and SNe Ia \cite{SNobs}, they all probe essentially the
same distance measure at different redshifts.
(The luminosity distance and the angular diameter distance
are directly related to each other. In FRW models, they can
be expressed in terms of the proper distance and
redshift. For the different distance measures in
the context of FRW models, see \cite{Hogg:1999, Cattoen}.)

The details of the expansion rate (or the distance scale)
are not well known except in specific models.
Model independent constraints are relatively weak.
The CMB is mostly sensitive to the angular diameter distance
to the last scattering surface, which is related to the
position of the peaks in the angular power
spectrum \cite{Efstathiou:1998, Mukhanov:2003}.
Type Ia supernovae provide a more direct measure of the expansion rate,
but present data is not sufficiently accurate to give detailed
information. Quoted constraints on, for example,
the equation of state are often driven by the assumed
parametrisation, which can be responsible for misleading apparent
precision and artificially small confidence level contours.
For the importance of parametrisation, see
\cite{Cattoen, trans, Shapiro:2005, Gong:2006, Elgaroy:2006, Linder:2007};
an analysis of the current situation in terms of a piecewise
constant equation of state is given in \cite{Zunckel:2007}.
Additional systematic effects, such as metallicity
\cite{Podsiadlowski:2006}, changes in the treatment of
dust \cite{Conley:2007} or the light-curve fitting method
\cite{Seikel:2007} can further degrade the reliability of
the SN Ia measurements.
It is noteworthy that the quality of fit of the \LCDM model has
decreased with the introduction of each new SN Ia dataset up to
and including the 'gold' sample and the ESSENCE data \cite{Vishwakarma:2005}.
This may hint at inadequacy of the \LCDM description, or
it can be indicative of underestimated systematic errors.

Observational constraints from perturbations
are much weaker than those from background quantities.
The Integrated Sachs-Wolfe (ISW) effect has been detected
via cross-correlation of the CMB and matter tracers at different
redshifts \cite{Ho:2008}. This is interpreted as
time evolution of the gravitational potential, which may be
due to accelerating expansion or spatial curvature.
The amplitude is slightly larger than expected in the \LCDM model,
though not significantly so (around 2$\sigma$).
Other constraints from the evolution of perturbations
are rather weak \cite{lineargrowth}.
A measure which combines background evolution
and the evolution of perturbations is the number count
of clusters as a function of redshift: in an accelerating model
the number density should rise sharply with increasing redshift.
It has been earlier argued that this is not seen in the data,
and that the observations instead prefer deceleration \cite{Blanchard}.
However, it seems that limited understanding of gas physics prevents
drawing reliable conclusions from cluster counts at
present \cite{Ferramacho:2007}.

In the context of homogeneous and isotropic models, we
can with confidence say that the expansion
has accelerated within the last few billion years (though it
is not necessarily accelerating at present), and determine the
overall magnitude of the acceleration
\cite{Shapiro:2005, Gong:2006, Elgaroy:2006, Seikel:2007}.
In order for a model to agree with the observations, it is not
necessary to reproduce the expansion history of the \LCDM model
in detail, only to produce a roughly similar amount of acceleration
in the same era.

\para{Structure formation and the coincidence problem.}

While there is no evidence apart from the expansion rate
for modified gravity or negative pressure matter, we know
that the assumption of linearly perturbed homogeneity and
isotropy breaks down in the universe at late times. 
One has to evaluate the effect of non-linear structure
formation on the expansion rate (or the distance scale) before
concluding that it is necessary to change either gravity or the matter content.

It was suggested in \cite{Schwarz, Rasanen}
that inhomogeneities related to structure formation could be
responsible for accelerated expansion (the possibility had been
earlier touched upon in \cite{Buchert:2000, Tatekawa:2001}).
This was discussed more concretely in
\cite{Rasanen:2006a, Rasanen:2006b}
where it was demonstrated with a toy model how the formation of
non-linear structures can lead to average acceleration,
even when the expansion locally decelerates
(see also \cite{Kai:2006}). The physical
reason is simply that the fraction of volume in faster expanding
regions rises, so the average expansion rate can rise.
The bigger is the difference between the slower and faster
expanding regions, the more rapid is the change as the faster
regions take over. 
It was pointed out in \cite{Rasanen:2006a, Rasanen:2006b}
that the growth of non-linear structures involves a growing variance
in the expansion rate, and that structure formation has a preferred
time around 10--100 billion years, near the observed acceleration
era. The timescale emerges from the CDM power spectrum, essentially
from the time of matter-radiation equality encoded in the
change of slope of the CDM transfer function.
This could solve the coincidence problem of why the acceleration
has started recently in cosmic history, something that the
\LCDM model does not explain.

The link between the preferred time in structure formation
and the change of the expansion rate was only conjectured in
\cite{Rasanen:2006a, Rasanen:2006b}. The toy model with
acceleration involved only two regions instead of a realistic
ensemble of structures, and had no link to the preferred time.
In the present work, we remedy both shortcomings.
We first discuss the effect of structures on the
propagation of light and the expansion rate in section 2,
and set up the Buchert backreaction formalism in section 3.
In section 4 we present our model for a statistically
homogeneous and isotropic universe containing an evolving
ensemble of non-linear structures.
We calculate the average expansion rate and find that
it grows relative to the FRW value around
10--100 billion years, though the change is not rapid enough
to correspond to acceleration.
We explain the reason and consider how improve our treatment.
In section 5 we discuss some observational and theoretical
issues related to backreaction, and summarise our results.
This work is a follow-up to \cite{Rasanen:2006b}, and
more discussion of these topics can be found there.
However, we have repeated some material to make
the presentation self-contained.

\section{Clumpy spacetimes}

\subsection{Light propagation in a clumpy spacetime} \label{sec:prop}

\para{The overall geometry.}

Though the evidence for negative pressure matter or modified
gravity is often phrased in terms of the expansion rate,
observations of light just provide constraints on the distance
scale at different redshifts.
The conclusion that the observations imply accelerated expansion
has been established only in the context of linearly perturbed
FRW models, where the distance scale is directly related to
the expansion rate (and spatial curvature).
The real universe is not locally perturbatively near homogeneity
and isotropy at late times, but contains non-linear structures (we
refer to such a universe as clumpy), so the FRW results for
light propagation cannot be straightforwardly applied.

Since non-linear structures
affect the propagation of light, it might be possible to
explain the observations without recourse to accelerated expansion.
We will be interested in the possibility that non-linear
structures lead to actual acceleration. But even in this case,
one needs to study light propagation in a clumpy space
to be able to compare the model to observations.

In FRW models, in order to convert between measures of
distance and the expansion rate, one needs to know the spatial
curvature. From the volume and spatial curvature of each
hypersurface of proper time, the distances can be reconstructed.
In a general spacetime, this is not true, and the distance
scale does not necessarily correspond to the expansion rate
(see e.g. \cite{Enqvist:2006}).
We can divide the effect of structures on the passage of light
into an overall part, which can be expressed in terms of
the average geometry given by the scale factor (which measures
the volume of the spatial hypersurface) and the trace of the
spatial Ricci tensor (which is a measure of the curvature of
the hypersurface), and a part which cannot be expressed in
terms of the average geometry.

For example, in perturbed FRW models, the ISW and Rees-Sciama effects
are not expressible in terms of the scale factor and spatial curvature.
These effects are small both because the perturbations are small and
because there are cancellations present.
In particular, in the linearly perturbed Einstein-de Sitter universe
(the spatially flat FRW dust model), the ISW effect is
zero independent of the magnitude of the gravitational potential
(as long as the second order contribution can be neglected).
In models which include realistic, non-perturbative structures,
there is no obvious reason for such effects to be small.
Nevertheless, it is possible that this is the case in a universe
which is statistically homogeneous and isotropic, and that the
passage of light can (up to small corrections) be described in
terms of the overall geometry.

\para{Studies of light propagation.}

There is currently no derivation of light propagation
for the case of a clumpy universe that contains realistic
evolving structures like those that are actually present.
However, several models with realistic elements have been studied.
The effect of inhomogeneities on the passage of light was
first discussed by Zel'dovich in 1964 \cite{Zeldovich:1964}
(early literature on the topic also refers to an unpublished
colloquium by Feynman in the same year). This and the work which followed
\cite{Dashevskii, Bertotti:1966, Gunn:1967}
studied the passage of light in the case when the deviation
from the FRW models is in some sense small (see
\cite{Sasaki:1987, Futamase:1989, Kasai:1990, Claudel:2000, Barausse:2005, Bonvin, Vanderveld:2007}
for later studies of perturbed FRW universes).

The effect of non-linear perturbations on the
redshift and the luminosity distance was studied in 1969 in the
context of the Swiss cheese model \cite{Kantowski:1969}, where
sections of FRW universe are cut out and replaced with the
Schwarzschild solution of equivalent density. It was found
that corrections to the FRW results could be sizeable.

The approach introduced by Zel'dovich in \cite{Zeldovich:1964},
where light rays encounter only a fixed fraction of the mass of
the universe because of clumping, was used by Dyer and
Roeder in 1972 to calculate corrections to the luminosity
distance \cite{Dyer} (and is now known as the Dyer-Roeder formalism).
They too concluded that large effects are possible.
The formalism was generalised to include
a mass fraction of clumps which varies with redshift in
\cite{Linder}. For applications of the Dyer-Roeder formalism
and Swiss cheese models to observations of SNe Ia,
see \cite{SN, Mattsson:2007b}.

It was argued by Weinberg in 1976 that the effects of
clumpiness on the luminosity distance cancel, and the
FRW formula can be used to describe light propagation \cite{Weinberg:1976}.
The main argument is that the number of photons is
conserved and gravitational deflection conserves photon energy,
so the luminosity distance can be determined simply in terms
of the area of a sphere drawn around the emission point.
However, it was pointed out in \cite{Ellis:1998a} that this argument
assumes that the area is the same as in the FRW model, which
is, in fact, the issue under study. An exact counter-example
was provided in \cite{Mustapha:1997}, and further arguments
that the FRW result may not apply were given in \cite{Ellis:1998b}.

In addition to purely analytical work
\cite{Linder:1998, Lieu:2004a, Kibble:2004, Lieu:2004b},
light propagation has been studied using ray-tracing methods
\cite{Kasai:1990, Watanabe:1989, Holz:1997, Sugiura:1999, Brouzakis, Marra:2007}.
Ray-tracing studies disagree on the magnitude of the effect on the
passage of light, presumably due to different modelling
assumptions. Among the most realistic studies are modified Swiss cheese
models where the non-linear density concentrations reside in the walls
around the holes, instead of the center. Studies of realistically
sized structures have found the corrections to be small,
at the percent level \cite{Sugiura:1999, Brouzakis}.
In \cite{Sugiura:1999} the integrated effect of an ensemble of voids
was found to be proportional to void size divided by the horizon size.
A sizeable correction was found in \cite{Marra:2007}, but this is
consistent with the previous results, since the structures
studied in \cite{Marra:2007} are unrealistically large.
However, the amplitude of the effect has not been conclusively settled,
since none of the models have been completely realistic, in particular
with regard to the time evolution of structures and spatial curvature.

In analytical work, there is also no agreement as to the
whether the effect can be large. It was found in \cite{Kibble:2004}
that for an ensemble of uncorrelated, static clumps of matter
in a FRW universe, the corrections are small (see
\cite{Lieu:2004a, Lieu:2004b} for
discussion of the small effects which could be observable).
According to \cite{Ellis:1998a, Ellis:1998b}, large effects are
possible due to the formation of caustics; see also \cite{Linder:1998}.
(For studies along related lines, see the program of observational
cosmology \cite{obscos}.)

As summarised in \cite{Ellis:2005} (see also \cite{Ellis:1999}),
one important issue is that light propagation is in general
affected both by the Ricci tensor given by the local matter
distribution and the non-locally determined Weyl tensor.
In FRW models, the Weyl tensor is zero and the geometry of
spacetime is entirely determined in terms of the local matter distribution.
In contrast, in the real universe, light mostly travels in vacuum
where the Ricci tensor is zero, and the geometry is determined by
the Weyl tensor. This difference is related to the
role of shear of the null geodesics, generated by nearby matter.
In this sense, the propagation of light in the real universe
is the reverse of the FRW situation.
(See \cite{Bertotti:1966, Dyer:1981} for early discussion.)

In contrast to the studies of an ensemble of structures
which we know to exist in the universe,
there has recently been revived interest in the idea that
we would live near the centre of a single untypically large
spherical void ('Hubble bubble').
Studies of the passage of light in the spherically symmetric
Lema\^{\i}tre--Tolman--Bondi (LTB) model \cite{LTB}
(see \cite{Krasinski:1997} for a review)
show that if the local bubble is sufficiently large, its effect
on the passage of light could explain the observations of SNe Ia
(and possibly be consistent with the location of the CMB peaks).
For references on the 'Hubble bubble' and the passage
of light in the LTB model, see \cite{Rasanen:2006b};
for a review on explaining the observations with the
LTB model, see \cite{Enqvist:2007}.

The results for the passage of light are not conclusive,
and no models have properly dealt with the kind of
nested, evolving structures that are a central feature of
hierarchical structure formation.
The effects which have been found for realistic
models of randomly distributed small structures are small.
Large effects have been demonstrated only when the observer
occupies a special location, such as near the center of a large
spherical region.
It therefore seems plausible that those effects of clumpiness on the passage
of light which have been studied are small if the matter distribution
is statistically homogeneous and isotropic to a sufficient level
and the observer does not occupy a special location.
(The small size of structures relative to the observed volume
can be considered part of homogeneity and isotropy.)
However, most studies of light propagation have neglected the influence
of structures on the average expansion rate and the spatial curvature,
assuming that they evolve according to the FRW equations.
So the results should not be interpreted as evidence that
the effect of structures is small, only that it is plausible
that their effect can be considered in terms of the overall geometry.

Even if that is true, one still needs to derive the
proper distance as a function of redshift in terms of the
expansion rate and spatial curvature, which describe
the overall geometry. In general, the average
spatial curvature does not evolve like $a^{-2}$, where $a$ is
the scale factor, and simply inserting spatial curvature with
arbitrary time-dependence into the FRW relations does not
make sense \cite{Rasanen:2007}.
(Light propagation in terms of a scale factor which takes into account
the influence of structures, but with the spatial curvature evolving
like in the FRW case, was discussed in \cite{Paranjape:2006b}.)
We will only study the evolution of the overall geometry
as described by the scale factor and average spatial curvature,
and will not consider the propagation of light.

From the practical point of view, treatment of the passage of light
in terms of a scale factor (even without any spatial curvature)
has been very successful in accounting for the observations.
If the observed deviation from the Einstein-de Sitter model
was due to effects which cannot be expressed in terms
of the average geometry, one would not expect different
observations to agree so well with the simple scale factor
treatment. For example, in the 'Hubble bubble' models,
the density gradient which accounts for 
the luminosity distances to SNe Ia typically does not explain
the angular diameter distance to the
last scattering surface, or the scale of the baryon acoustic
oscillations \cite{Mattsson:2007b, Enqvist:2007}.

\subsection{The expansion rate in a clumpy spacetime}

\para{Inadequacy of the FRW description.}

Assuming that the influence of clumpiness on the 
passage of light can be encapsulated in the scale factor
and the spatial curvature scalar, we are left with the question of
how non-linear structures affect the expansion rate and the spatial
curvature.

The linearly perturbed FRW equations do not describe the local
evolution of the real universe. Considering the linearly
perturbed Einstein-de Sitter metric (and neglecting the
decaying mode), the local expansion rate measured by a comoving
observer is $\theta=3H - \delta H$,
where $H$ is the background Hubble parameter in terms of the proper
time and $\delta\propto a$ is the density contrast \cite{Rasanen}.
When $\delta$ becomes of order unity, these relations
fail to describe the real behaviour.
In the case of underdensities, this is obvious, since the density
contrast cannot decrease below $-1$.
For typical spherical overdense regions, the difference between
the linear and non-linear evolution is well-known from the
spherical collapse model \cite{Gunn:1972} (see
\cite{Padmanabhan:1993, Liddle:2000, Sheth} for reviews).

It has been argued that even if there are large deviations locally,
the statistical homogeneity and isotropy of the universe
implies that the average evolution follows the FRW equations.
However, a space which is statistically homogeneous and isotropic
does not in general evolve on average like a space which is
exactly homogeneous and isotropic.
Exact homogeneity and isotropy leads to the FRW equations.
Statistical homogeneity and isotropy simply supports the
assumption that one can make sense of the observations in
terms of the overall geometry.
This does not imply that the scale factor follows the FRW equations,
essentially because the spatial curvature in a clumpy spacetime
in general evolves differently from the FRW case.
(A test of whether the metric has the FRW form
was proposed in \cite{Clarkson:2007}, based on the specific
FRW evolution of spatial curvature.)

The influence of inhomogeneity and/or anisotropy on the expansion
rate is known as backreaction, and has been discussed in a number
of papers
\cite{fitting, Buchert:1995, Buchert:1999, Sicka, Carfora, Buchert:2001, Schwarz, Rasanen, Kolb:2004a, Rasanen:2006a, Rasanen:2006b};
see \cite{Rasanen:2006b, Ellis:2005, Buchert:2007} for reviews
and more references. (See \cite{Woodard, Unruh, Brandenberger:2002, Geshnizjani:2002, Geshnizjani:2003}
for discussion of backreaction during inflation.)

\para{Statistical homogeneity and isotropy.}

The early universe was highly homogeneous and isotropic,
so the FRW equations provide a good approximation of
the average evolution, and inhomogeneities and anisotropies
are described by linear perturbations around the FRW model.
Each hypersurface of constant proper time practically corresponds
to a single expansion rate (with tiny variations).
As perturbations grow and structures form, a given moment
of time no longer corresponds to a single expansion rate.
Instead, the hypersurface of constant proper time contains
regions in different stages of expansion, some of them
collapsing or static.
On sufficiently large scales, the spatial distribution of expansion
rates at each moment is statistically homogeneous and
isotropic, and we can meaningfully call the local expansion
rate averaged over all regions the expansion rate at that time.

In order for the average expansion rate at a given time to be a useful
quantity for describing the passage of light through
structures, it is important that the expansion rate changes
slowly compared to the time it takes for light to cross typical
structures. Light has to have time to pass through different
regions and sample the distribution of expansion rates before
it changes appreciably. Otherwise, it does not make sense to use
an expansion rate averaged over many regions: one would have
to describe the passage of light through the individual regions.

This condition is well satisfied in the real universe. For typical
supersymmetric weakly interacting dark matter, the first structures
which form around a redshift of $z\sim 40-60$ have sizes of the
order $10^{-8} H^{-1}$ \cite{SUSYCDM}, and typical largest structures
today have sizes around 10 \mpc{} $\approx 3\times 10^{-3} H^{-1}$
(where $h$ parametrises the present-day
Hubble rate, $H_0=100h$ km/s/Mpc).
A more important quantity is the homogeneity scale,
by which we mean the scale where
averages converge to their asymptotic value.
The fractal dimension of the set of galaxies around
us indicates a homogeneity scale of 70--100 \mpc
\cite{Hogg:2004, Pietronero}, while studies of morphology
suggest that it is at least 200 \mpc
\cite{morphology}\footnote{The fact that there seems to be a
significant contribution to our motion due to the Shapley Supercluster
at a distance of 130--180 \mpc \cite{SSC} suggests that the homogeneity
scale is on the larger side. Alternatively, our location in the universe
may be rather untypical.}.
In either case, the homogeneity scale is $\approx 10^{-2} H^{-1}$,
so light rays pass through several representative regions of the
universe in one Hubble time, which is the timescale for
significant change in the expansion rate.

A related concern about the applicability of the averaged
expansion rate is the question of over which scale the averaging
is done. To get a representative sample, the averaging scale should
be at least as large as the homogeneity scale.
Because of statistical homogeneity and isotropy,
it should not matter if the averaging scale is larger than this.
In practical terms, most cosmological observations of the expansion
rate probe distances larger than the homogeneity scale, apart
from SNe Ia at small redshifts.
For a statistically homogeneous and isotropic
distribution of structures, varying the averaging scale
as discussed in \cite{Mattsson:2007a} should be relevant
only for local observations, where the average scale
factor treatment may be anyway problematic.
(The observed large angle anomalies in the CMB
might be related to local departures from the scale factor
description; see \cite{Rasanen:2006b} for discussion.)

The issue of statistical homogeneity and isotropy is related to
the choice of the hypersurface of averaging. The scale factor describes the
volume of the hypersurface of proper time, but why take that hypersurface?
Observations are organised on hypersurfaces of constant redshift, not
proper time. However, if we can identify the redshift with a scale
factor in the usual way, $1+z=a(t)^{-1}$, then we have a one-to-one
correspondence between the redshift $z$ and proper time $t$, and the
hypersurfaces of constant proper time and constant redshift agree.

The question of the choice of hypersurface is also present in
FRW models, and the hypersurface of constant proper
time is selected because it is the hypersurface of homogeneity
and isotropy. In realistic models, there is no such exact symmetry,
but one still expects the hypersurface of constant proper time 
to agree with the hypersurface of statistical homogeneity and isotropy,
and the argument for choosing this hypersurface is the same
as in the FRW case.
The evolution of structures proceeds according to proper time,
so were one to tilt the hypersurface, different parts would
contain structures which have evolved for different amounts of
time, breaking statistical homogeneity and isotropy.

The notion of statistical homogeneity
and isotropy with a given homogeneity scale is easy to grasp
intuitively: the universe consists of statistically identical
boxes of a size given by the homogeneity scale. When evaluating any average
physical quantities inside one box, the result should be independent
of the location and orientation of the box, up to statistical
fluctuations. This does not
imply that the average quantities must be those of a model
with exact homogeneity and isotropy. In terms of the Buchert
equations discussed below, the backreaction variable $\sQ$
saturates at the homogeneity scale, but not necessarily to zero.
This view neglects long-range correlations, and in practice
the degree of statistical homogeneity and isotropy increases as the
size of the boxes grows, up to the level $10^{-5}$ given by the
primordial perturbations. The concept of
statistical homogeneity and isotropy in a spacetime
with non-linear structures should be made more rigorous.
Studies of the choice of hypersurface in backreaction
\cite{Geshnizjani:2002, Geshnizjani:2003, Rasanen:2004}
have not touched on this issue.

\section{The Buchert formalism} \label{sec:Buchert}

\subsection{The local equations}

\para{The dust assumption.}

We assume that the matter content of the universe can be described as
dust, i.e. an ideal fluid with zero pressure. This is not true
on small scales (for example, the matter in the solar system cannot be
treated as a pressureless ideal fluid), so there is an implicit
averaging involved, distinct from the large-scale averaging we
are going to discuss. Essentially, the assumption is that
on sufficiently large scales we can model a complex system of
discrete small-scale structures as a continuum of infinitely fine
particles. The validity of this approximation has been studied
in the work on discreteness and the fluid approximation in N-body
simulations \cite{discrete}; indirect support can be found in the
'renormalisability' of Newtonian gravity in \cite{renorm} (see also 
\cite{Crocce:2007}).

\para{The Einstein equation.}

For dust, the Einstein equation reads
(taking the cosmological constant to be zero)
\bea \label{Einstein}
  G_{\a\b} &=& 8 \pi \GN T_{\a\b} \el
  &=& 8 \pi \GN \rho\, u_{\a} u_{\b} \ ,
\eea

\noindent where $G_{\a\b}$ is the Einstein tensor, $\GN$ is
Newton's constant, $T_{\a\b}$ is the energy--momentum tensor,
$\rho$ is the energy density and $u^{\a}$
is the velocity of observers comoving with the dust.

Following the covariant coordinate-independent approach, the Einstein
equation \re{Einstein} can be decomposed into scalar, vector and
tensor parts. This is a local decomposition with respect
to general coordinate transformations, not a global decomposition
in terms of the symmetry of a background, as in perturbed
FRW spacetimes. (We are not assuming any symmetries, or
making a division into background and perturbations.)
For reviews of the covariant approach, see
\cite{Ehlers:1961, Ellis:1971, Ellis:1998c, Tsagas:2007}.

We want to discuss averages. Since vector and tensor quantities
cannot be straightforwardly averaged\footnote{Though see the
work on the Ricci flow \cite{Carfora} and the 'macroscopic
gravity' formalism \cite{Zalaletdinov}. On the relation
of the latter to the averaging scheme used here, see \cite{Paranjape}.},
we will consider the scalar part of the Einstein equation.
The Einstein equation \re{Einstein} has two scalar components, and
the covariant conservation law yields a third equation
\cite{Raychaudhuri:1955, Ehlers:1961, Ellis:1971, Ellis:1990}
(see e.g. \cite{Buchert:1999} for the full system of equations):
\bea
  \label{Rayloc} \dot{\theta} + \frac{1}{3} \theta^2 &=& - 4 \pi \GN \rho - 2 \sigma^2 + 2 \omega^2 \\
  \label{Hamloc} \frac{1}{3} \theta^2 &=& 8 \pi \GN \rho - \frac{1}{2} \sR + \sigma^2 - \omega^2 \\
  \label{consloc} \rhodot + \theta\rho &=& 0 \ ,
\eea

\noindent where a dot stands for derivative with respect to
proper time $t$ measured by observers comoving with the dust,
$\theta$ is the expansion rate of the local volume element,
$\sigma^2=\ha\sigma^{\a\b}\sigma_{\a\b}\geq0$
is the scalar built from the shear tensor $\sigma_{\a\b}$,
$\omega^2=\ha\omega^{\a\b}\omega_{\a\b}\geq0$
is the scalar built from the vorticity tensor $\omega_{\a\b}$,
and $\sR$ is the Ricci scalar on the tangent
space orthogonal to the fluid flow.
The acceleration equation \re{Rayloc} is known as the Raychaudhuri
equation, and \re{Hamloc} is the Hamiltonian constraint.
The equations are exact, and valid for arbitrary large variations
in density, expansion rate and other physical quantities.

\para{Vorticity.}

When the vorticity is zero, the tangent spaces orthogonal to
the fluid flow form spatial hypersurfaces which provide
a foliation that fills the spacetime exactly once. These
flow-orthogonal hypersurfaces are also hypersurfaces of constant proper
time of comoving observers. If the vorticity is non-zero, no such
hypersurfaces exist. (See \cite{Ellis:1990} for discussion of the
Ricci scalar of the tangent spaces in the case of non-zero vorticity.)

The covariant approach deals directly with physical quantities, and
there is no need to introduce coordinates.
Nevertheless, it can be shown by explicit construction that if
we wish to choose coordinates, it is always possible to take
$g_{00}=-1$, and simultaneously obtaining $g_{0i}=0$ is possible
if and only if the vorticity is zero (here $0$ stands for time
and $i$ for the spatial directions) \cite{Ellis:1967}.
For irrotational dust, the synchronous
metric can be adopted locally without any loss of generality.
In the covariant formulation, there are no artifacts related
to the choice of coordinates, since there are no coordinates,
and all variables are physical observables. However, there would be
no problem (when the vorticity is zero) in using the synchronous
comoving coordinates. The results for physical quantities of course
do not depend on the coordinate system. (For example, in
perturbative backreaction calculations one obtains the same results
in the longitudinal and comoving synchronous gauges \cite{Kolb:2004a}.)

We take the vorticity to be zero. For dust,
if vorticity is initially zero, it will remain zero.
Furthermore, vorticity in linearly perturbed
FRW models corresponds to vector perturbations,
and it decays with expansion (unlike shear).
However, the description of matter as a
pressureless ideal fluid will break down on small scales when
non-linear structures form, due to shell-crossing.
After any vorticity is generated in gravitational
collapse \cite{DelPopolo:2001}, it will be
amplified by the collapse, and will formally diverge at the same
time as the density \cite{Buchert:1992}.
A static dust structure, with $\theta=0$, is possible only
if large amounts of vorticity are present, as \re{Rayloc} shows:
vorticity has to balance the contribution of both
the energy density and the shear on the right-hand side.
For stabilised dust structures, vorticity is the
dominant contribution. (As the approximation
of treating matter as dust breaks down, effects such as
velocity dispersion and pressure can be also important
in the stabilisation \cite{Buchert:2005a}.)

Nevertheless, just as we assume that the small-scale breakdown of the
description of matter as dust is not important, we assume
that the vorticity inevitably present at small scales can be
neglected in discussing the overall large-scale evolution.
(This assumption is also involved in the usual perturbed FRW treatment,
with rotationless ideal fluids and a well-defined cosmic time.)
Vorticity is probably only relevant in stabilising
structures. Since we do not need to consider the details of
stabilisation in our calculation, we do not expect the
assumption of zero vorticity to be important.

\subsection{The average equations} \label{sec:average}

\paragraph{Defining the average.}

We follow the formalism introduced by Buchert in
\cite{Buchert:1995, Buchert:1999, Buchert:2001}.
The spatial average of a quantity is its integral over the
hypersurface of constant proper time $t$, divided by the volume
of the hypersurface
\bea \label{av}
  \av{f}(t) \equiv \frac{ \int_t \epsilon f }{ \int_t \epsilon } \ ,
\eea

\noindent where $\epsilon_{\a\b\c}=\eta_{\a\b\c\d} u^\d$
is the volume element on the hypersurface of proper time,
$\eta_{\a\b\c\d}$ being the spacetime volume element.

The scale factor is defined simply as the volume of the
hypersurface of constant proper time to power $1/3$
\bea  \label{a}
  a(t) \equiv \left( \frac{ \int_t \epsilon}{ \int_{t_0} \epsilon} \right)^{\frac{1}{3}}  \ ,
\eea

\noindent where $a$ has been normalised to unity at time $t_0$,
which we take to be today. As $\theta$ is the volume expansion
rate, this definition of $a$ is equivalent to $3\adot/a\equiv\av{\theta}$.
We will also use the notation $H\equiv\adot/a$.

\paragraph{The Buchert equations.}

Let us take the average of the equations \re{Rayloc}--\re{consloc}.
The resulting Buchert equations are \cite{Buchert:1999}:
\bea
  \label{Ray} 3 \frac{\addot}{a} &=& - 4 \pi \GN \av{\rho} + \sQ \\
  \label{Ham} 3 \HH &=& 8 \pi \GN \av{\rho} - \frac{1}{2}\av{\sR} - \frac{1}{2}\sQ \\
  \label{cons} && \pat_t \av{\rho} + 3 \H \av{\rho} = 0 \ ,
\eea

\noindent where the backreaction variable $\sQ$ contains the effect
of inhomogeneity and anisotropy:
\bea \label{Q}
  \sQ \equiv \frac{2}{3}\left( \av{\theta^2} - \av{\theta}^2 \right) - 2 \av{\sigma^2} \ .
\eea

\noindent The integrability condition for the average Raychaudhuri
equation \re{Ray} and the average Hamiltonian constraint \re{Ham} is
\bea \label{int}
  \pat_t {\av{\sR}} + 2 \H \av{\sR} = - \pat_t \sQ - 6 \H \sQ \ ,
\eea

The Buchert equations \re{Ray}--\re{cons} are exact for the averages
when matter consists of irrotational dust. (The corresponding equations
in the Newtonian case were derived in \cite{Buchert:1995}, and the
case with non-zero pressure was considered in \cite{Buchert:2001}.)
The variance of the expansion rate in $\sQ$ is a new term compared
to the local equations. It has no counterpart in the local dynamics,
and may be called emergent in the sense that it is purely a property
of the averaged system. If the variance is large enough, the average
expansion rate can accelerate, even though the local expansion rate
decelerates everywhere. (In physical terms, the average expansion
rate can grow because the volume occupied by faster expanding regions
rises.)

\paragraph{The density parameters.}

As in the case of FRW models, we can parametrise the different
contributions to the expansion rate with relative densities.
Dividing \re{Ray} and \re{Ham} by $3 H^2$, we have \cite{Buchert:1999, Buchert:2003}
\bea \label{omegas}
  \label{q} q &\equiv& - \frac{1}{H^2} \frac{\addot}{a} = \frac{1}{2} \Om + 2 \OQ \\
  \label{Omegas} 1 &=& \Om + \OR + \OQ \ ,
\eea

\noindent where $\Om\equiv 8\pi G_N \av{\rho}/(3 H^2)$,
$\OR\equiv-\av{\sR}/(6 H^2)$ and $\OQ\equiv-\sQ/(6 H^2)$ are
the density parameters of matter, spatial curvature and
the backreaction variable, respectively. As seen from the
definition of $\sQ$ in \re{Q}, the backreaction density parameter
is just minus the relative variance of the expansion rate, plus the
contribution of shear:
$\OQ = - ( \av{\theta^2} - \av{\theta}^2 )/\av{\theta}^2 + 3 \av{\sigma^2}/\av{\theta}^2$.

The backreaction density $\OQ$ can have either sign, and can become arbitrarily
negative. Correspondingly, $q$ can cross the value $-1$, and the contribution
$\OR + \OQ$ can change sign. These features are not captured
in the parametrisation of backreaction in terms of a scalar
field, the 'morphon' \cite{morphon}.

\subsection{First integrals of the Buchert equations} \label{sec:first}

\paragraph{First integral in terms of $\sQ$.}

The scalar parts of the Einstein equation \re{Rayloc} and \re{Hamloc}
together with the conservation law \re{consloc} do not form a closed
system on their own. (In particular, the propagation equations for
the shear and vorticity tensors cannot be reduced to scalars.)
There are four unknowns ($\theta, \rho, \sigma^2-\omega^2, \sR$)
and three equations. The Buchert equations \re{Ray}--\re{cons} are
similarly underdetermined, with four unknowns
($a, \av{\rho}, \sQ, \avR$) and three equations.

If we give as extra input the evolution of one of the variables
(or a relation between the variables), the average system is
completely determined.
In particular, we can find the average expansion rate
from the average spatial curvature $\av{\sR}$
or the combination of variance and shear given in $\sQ$.
We can make this explicit with the first
integral of the Buchert equations \re{Ray}--\re{cons}.

Taking into account that the conservation of mass \re{cons}
implies $\av{\rho}\propto a^{-3}$, we can integrate \re{Ray} to obtain
\bea \label{firstQ}
  3 H^2 &=& 8 \pi \GN \frac{\av{\rho_0}}{a^3} - 3 \frac{K}{a^2} + \frac{2}{a^2}\int^a \frac{\rmd a'}{a'}\, a'{}^2\sQ \ ,
\eea

\noindent where $K$ is an integration constant related to
the spatial curvature and $\av{\rho_0}$ is the average energy
density at time $t_0$.
When backreaction vanishes, we recover the FRW Hubble relation,
and the spatial curvature is $6 K a^{-2}$, as usual. When the
spacetime is clumpy and  backreaction is present, $\sQ\neq0$,
the average spatial curvature evolves non-trivially:
\bea \label{RfirstQ}
  \frac{1}{2} \av{\sR} &=& 3 \frac{K}{a^2} - \ha\sQ - \frac{2}{a^2}\int^a \frac{\rmd a'}{a'}\, a'{}^2\sQ \ .
\eea

\paragraph{First integral in terms of $\avR$.}

It is also instructive to write the first integral
in terms of the spatial curvature. Expressing things
in terms of $\avR$, we have from \re{Ray}--\re{cons},
\bea \label{firstR}
  3 H^2 = 8 \pi \GN \frac{\av{\rho_0}}{a^3} + \frac{C}{a^6} - \frac{2}{a^6} \int^a \frac{\rmd a'}{a'}\, a'{}^6 \avR \ ,
\eea

\noindent where $C$ is an integration constant. The backreaction variable is
\bea \label{QfirstR}
  \sQ = - 2 \frac{C}{a^6} - \av{\sR} + \frac{4}{a^6} \int^a \frac{\rmd a'}{a'}\, a'{}^6 \avR \ .
\eea

When the average spatial curvature evolves like $a^{-2}$, the $C$
term gives the backreaction variable $\sQ$, which in accordance
with the integrability constraint \re{int} then evolves like $a^{-6}$.
We can make this more explicit by decomposing the average
spatial curvature as $\avR=6 K a^{-2} + \Delta\avR$.
The backreaction variable is\footnote{In
\cite{Gruzinov:2006} it was argued on the basis of a 2+1-dimensional
study that the effect of backreaction is small. However, since the Ricci
scalar in two dimensions is a topological invariant, the average
two-dimensional spatial curvature is necessarily proportional to
$a^{-2}$, and the evolution of $\sQ$ is trivial.
Therefore the situation in 2+1 dimensions is not representative
of the 3+1-dimensional case, where spatial curvature
is dynamical. There were also two other arguments presented in
\cite{Gruzinov:2006}, one based on second order perturbation
theory and the other on a solution with parallel walls.
The first calculation is incorrect, because it assumes that
the background energy density is the same as the average energy
density, which is not true beyond the linear level. In the
second case, variations in the local expansion rate are small,
so it is irrelevant for the real universe.}
\bea
  \sQ = - 2 \frac{C}{a^6} + \frac{4}{a^6} \int^a \frac{\rmd a'}{a'}\, a'{}^6 \Delta \avR \ ,
\eea

\noindent and the Hubble equation \re{firstR} reads
\bea \label{firstRb}
  3 H^2 = 8 \pi \GN \frac{\av{\rho_0}}{a^3} - 3 \frac{K}{a^2} + \frac{C}{a^6} - \frac{2}{a^6} \int^a \frac{\rmd a'}{a'}\, a'{}^6 \Delta\avR \ .
\eea

In general, underdense regions are negatively curved
and expand faster than the average, while overdense regions
are positively curved and expand slower. The evolving
distribution of these regions determines the average spatial curvature.
At late times, one would expect the faster expanding regions to dominate
the volume, and the spatial curvature to be negative. The intertwining
of the expansion rate and the spatial curvature is quantified in \re{firstRb}.
While $\sQ$ is non-local, $\av{\sR}$ is simply the average of a
local quantity; they are interchangeable
via the integrability condition \re{int}.
Understanding backreaction in terms of an average over the
local spatial curvature will be useful when we discuss
the importance of non-Newtonian aspects of gravity in backreaction
in section \ref{sec:Newton}.

We need only know the evolution of the variance of the expansion
rate minus the shear in order to reconstruct the complete
expansion rate (up to a constant related to the spatial
curvature). There is no need to specify the full metric, only
some statistics.
Indeed, it is unfeasible to write down a metric that would
be a realistic description of the structures in the universe.
Exact solutions such as the spherically symmetric LTB model are
useful for illuminating specific aspects of backreaction
such as the choice of hypersurface \cite{Rasanen:2004}
and demonstrating acceleration unambiguously
\cite{Kai:2006, Chuang:2005, Paranjape:2006a}.
However, exact solutions, even ones with no symmetry such as the
Szekeres model \cite{Szkeres}, are limited in scope.
While more complicated than the LTB model, they are not much closer
to a realistic picture of the universe with its hierarchical
layers of individually complex but statistically homogeneous
and isotropic structures.
In order to evaluate the importance of backreaction in the real
universe, we need statistical knowledge about complex
configurations of dust, not exact information about simplified models.

\section{Backreaction with the peak model}

\subsection{Setting up the model} \label{sec:setup}

\para{The universe divided into regions.}

We want to have a model of the universe that does not violate
large-scale statistical homogeneity and isotropy, and includes
a realistic ensemble of evolving structures. We will treat
each structure as an isolated region, and take the
nested nature of hierarchical structure formation into
account in the evolution of the number density of the regions.

We divide the hypersurface of constant proper time into
non-overlapping regions labelled by $\delta$.
The fraction of proper volume occupied by each region is
$v_\d(t)\equiv\int_{t,\d} \epsilon /\int_t \epsilon$,
where $\int_{t,\d}$ is the integral over region $\d$ at time $t$.
This division is completely general.
We now take the regions to be isolated and spherically
symmetric.
This is an extension of the two-region toy model studied
in \cite{Rasanen:2006a, Rasanen:2006b} to cover a realistic
distribution of structures.
An ensemble of only spherically symmetric regions cannot
fully cover the hypersurface of proper time (and does
not form a connected space), and the model has to be
understood in a statistical sense. (We come back to this
point in \sec{sec:Newton}.)

We treat the spherical regions with Newtonian gravity.
Their average evolution is then given by the spherical
collapse model (and its underdense equivalent), according
to which they evolve like the corresponding FRW universe
\cite{Gunn:1972} (see
\cite{Padmanabhan:1993, Liddle:2000, Sheth} for reviews).
In terms of the Buchert equations, this follows
from the result that in Newtonian gravity, $\sQ$ vanishes
for spherical symmetry, so the Buchert equations reduce to the
FRW equations \cite{Sicka, Buchert:2000}.
Expansion of a region with positive density contrast slows
down more as the perturbation grows, until it turns
around and collapses, finally stabilising at a finite size
and density. Inside a region with negative density contrast,
expansion decelerates less as the region becomes emptier.

We take the individual regions of the ensemble to represent
structures at a definite state of expansion or collapse.
In other words, the label $\d$ corresponds to a
value of the average expansion rate in a region.
In the spherical collapse model, the expansion rate
is in one-to-one correspondence with the
linear density contrast (hence the letter $\delta$),
which is the density contrast that the region would have
relative to an Einstein-de Sitter universe if its evolution
had continued as in the linear regime.

The average expansion rate is
\bea 
  \frac{1}{3} \av{\theta} = H(t) &=& \int_{-\infty}^{\infty} \rmd\delta\, v_\d(t) H_\d(t) \el
  &=& \frac{ \int_{-\infty}^{\infty} \rmd\delta\, s_\d f(\d,t) H_\d(t)}{ \int_{-\infty}^{\infty} \rmd\delta\, s_\d f(\d,t) }
\ ,
\eea
\noindent The fraction of proper volume in structures with linear
density contrast $\d$ at time $t$ has been decomposed into two parts,
$v_\d(t)=s_\d f(\d,t)/( \int_{-\infty}^{\infty} \rmd\delta\, s_\d f(\d,t))$.
(Note that the linear density contrast can be arbitrarily negative,
unlike the real density contrast, which is bounded from below by $-1$.)

The first term $s_\d\equiv a_\d(t)^3/\afrw(t)^3$ is the volume of a
region with linear density contrast $\delta$ relative to
the volume of the Einstein-de Sitter universe, where
$\afrw\propto t^{2/3}$
(we will also use the notation $\Hfrw=2/(3t)$).
It is due to the difference in the expansion rate
between the different regions: more underdense regions
expand faster and have therefore grown larger.
(The scale factor $\afrw$ has been introduced out
of convenience. It appears both in the numerator and the
denominator and does not depend on $\delta$, so it does not
affect the average.)

The second term $f(\d,t)$ is the fraction of the initial
volume in regions with linear density contrast $\d$.
This is updated to the present volume by the first term.
If we completely ignored interactions between regions, 
$f(\d,t)$ would be directly given by the linear spectrum
of perturbations. However, we want to take into account the
nested nature of cosmological perturbations, and
the merger of structures into larger entities. As overdense
regions collapse and stabilise, they form larger structures
which in turn slow down and collapse, and underdense regions
have similar hierarchical evolution.
We will include this feature using the
peak model of structure formation.

The premise of the peak model of structure formation is that
structures are identified with maxima of the linear, Gaussian
density field, smoothed on some scale $R$ \cite{Bardeen:1986}.
In the original application, all peaks above a fixed density
threshold were considered to be stabilised non-linear structures.
We will use the peak number density as a function of density
contrast as the number density of isolated regions having 
that linear density contrast.
(Since the density field is Gaussian, the distribution of underdense
troughs is the same as the distribution of overdense peaks.)
The number density can be converted into the fraction of mass.
Since the early universe was very smooth, this is also the fraction
of the initial volume which has ended up in structures of a given
linear density contrast, in other words our $f(\d,t)$.

\para{The backreaction variable.}

Our treatment will give us the average expansion rate directly,
without needing to go via the backreaction variable $\sQ$.
However, interpreting the expansion rate in terms of the Buchert
equations \re{Ray}--\re{cons} and the density parameters
\re{omegas} will help to understand its evolution.
The backreaction variable for the model is
\bea \label{Qpeak}
  \sQ &=& \frac{2}{3} \left( \av{\theta^2} - \av{\theta}^2 \right) - 2 \av{\sigma^2} \el
  &=& \frac{2}{3} \left( \int_{-\infty}^{\infty}\rmd\d\,  v_\d \av{\theta^2}_\d - \left( \int_{-\infty}^{\infty}\rmd\d\,  v_\d \av{\theta}_\d \right)^2 \right) - 2 \int_{-\infty}^{\infty}\rmd\d\,  v_\d \av{\sigma^2}_\d \el
  &=& \frac{2}{3} \left( \int_{-\infty}^{\infty}\rmd\d\,  v_\d \av{\theta}_\d^2 - \left( \int_{-\infty}^{\infty}\rmd\d\,  v_\d \av{\theta}_\d \right)^2 \right) \el
  && + \frac{2}{3} \left( \int_{-\infty}^{\infty}\rmd\d\,  v_\d \av{\theta^2}_\d - \int_{-\infty}^{\infty}\rmd\d\,  v_\d \av{\theta}_\d^2 \right) - 2 \int_{-\infty}^{\infty}\rmd\d\,  v_\d \av{\sigma^2}_\d \el
  &=& 6 \left( \int_{-\infty}^{\infty}\rmd\d\,  v_\d H_\d^2 - \left( \int_{-\infty}^{\infty}\rmd\d\,  v_\d H_\d \right)^2 \right) + \int_{-\infty}^{\infty}\rmd\d\,  v_\d \sQ_\d \ ,
\eea

\noindent where $\av{}_\d$ is the average over region $\d$.
The total backreaction variable $\sQ$ is not the sum of the
regional backreaction variables, due to the non-local variance term.
Since we consider spherical regions using Newtonian gravity,
we have $\sQ_\d=0$ \cite{Sicka, Buchert:2000}, and the overall
backreaction variable can be calculated from the regional average
expansion rates.

\para{The spherical collapse/expansion model.}

The expansion rates of the regions follow the spherical
collapse model (and its underdense equivalent), which we
summarise here (see \cite{Sheth} for another useful summary).

For overdense regions we have
\bea \label{SCM}
  H_{\d^{+}}t = \frac{ \sin\phi (\phi-\sin\phi) }{ (1-\cos\phi)^2 } \el
  s_{\d^{+}} \equiv \frac{a^3_{\d^{+}}}{\afrw^3} = \frac{ 2 (1-\cos\phi)^3 }{ 9 (\phi-\sin\phi)^2 } \el
  \av{\sR}_{\d^{+}} t^2 = 6 \frac{ (\phi-\sin\phi)^2 }{ (1-\cos\phi)^2 } \el
  \d^{+} = \frac{3}{20} 6^{2/3} (\phi-\sin\phi)^{2/3} \ ,
\eea

\noindent where $\phi$ is the development angle which
runs from 0 to $2\pi$ and $\delta^+$ is the linear density contrast.
The structure collapses to a singularity at $2\pi$,
so we take the expansion rate to stabilise discontinuously
to zero at $\phi=2\pi$.
We set the volume, expansion rate and spatial curvature of
regions with a linear density contrast larger than
$\d^+(2\pi)\approx1.7$ to zero.
It would be more realistic to set the volume to a finite
constant and the spatial curvature to some constant proportional
to the stabilised energy density, but this makes little
difference for our purposes.
In our calculation, each region is weighted by its relative
volume, so the contribution of collapsing regions goes to zero
at the collapse, and the details of stabilisation do not matter.
(For example, the results would be unchanged if we stopped
the evolution at $\theta=3\pi/2$ instead of $2\pi$, as is sometimes done.)
While the expansion rate is discontinuous at the stabilisation,
the volume-weighted expansion rate $s_{\d^{+}} H_{\d^{+}}t$ is continuous.
The combinations $H_{\d^{+}}t$, $s_{\d^{+}}$ and $\av{\sR}_{\d^{+}} t^2$ do not
depend explicitly on $t$, they are functions of $\d^{+}$ alone.

Correspondingly, for underdense regions we have
\bea \label{SEM}
  H_{\d^{-}}t = \frac{ \sinh\phit (\sinh\phit-\phit) }{ (\cosh\phit-1)^2 } \el
  s_{\d^{-}} \equiv \frac{a^3_{\d^{-}}}{\afrw^3} = \frac{ 2 (\cosh\phit-1)^3 }{ 9 (\sinh\phit-\phit)^2 } \el
  \av{\sR}_{\d^{-}} t^2 = - 6 \frac{ (\sinh\phit-\phit)^2 }{ (\cosh\phit-1)^2 } \el
  \d^{-} = - \frac{3}{20} 6^{2/3} (\sinh\phit-\phit)^{2/3} \ ,
\eea

\noindent where the development angle $\phit$ runs from $0$ to
$\infty$, corresponding to increasing time and decreasing density
contrast.

Regions with zero density contrast have $\Hfrw t=2/3$, $s_{\d^0}=1$
and zero spatial curvature.

\paragraph{Peak statistics.}

For a linear density field with Gaussian, statistically
homogeneous and isotropic perturbations smoothed on scale $R$, the number
density of peaks (troughs) of height (depth) $\nu\equiv\delta/\sigma_0(t,R)$
is \cite{Bardeen:1986}
\bea \label{n}
  n(\nu,R) = e^{-\frac{1}{2} \nu^2} \frac{1}{(2\pi)^2 R_*(R)^3} G(\nu,\gamma(R)) \ ,
\eea

\noindent where the function $G(\nu,\gamma(R))$ is
\bea \label{G}
  G(\nu,\gamma) = \int_0^{\infty} \rmd x F(x) \frac{1}{\sqrt{2\pi (1-\gamma^2)}} e^{\frac{-(x-\gamma|\nu|)^2}{2 (1-\gamma^2)}} \ ,
\eea

\noindent with

\bea \label{F}
  F(x) &=& \frac{x^3 - 3 x}{2} \left\{ \erf\left[\left(\frac{5}{2}\right)^{\ha} x\right] + \erf\left[\left(\frac{5}{2}\right)^{\ha} \frac{x}{2}\right] \right\} \el
  && + \left( \frac{2}{5\pi} \right)^{\ha} \left[ \left( \frac{31 x^2}{4} + \frac{8}{5} \right) e^{-\frac{5 x^2}{8}} +\left( \frac{x^2}{2} - \frac{8}{5} \right) e^{-\frac{5 x^2}{2}} \right] \ .
\eea

\noindent The functions $\gamma(R)$ and $R_*(R)$ are defined as (note
that the explicit time-dependence in $\sigma^2_j(t,R)$ cancels out)
\bea \label{gammaRstar}
  \gamma(R) &\equiv& \frac{\sigma^2_1(t,R)}{\sqrt{\sigma^2_0(t,R)\sigma^2_2(t,R)}} \el
  R_*(R) &\equiv& \sqrt{ 3 \frac{\sigma^2_1(t,R)}{\sigma^2_2(t,R)} } \ ,
\eea

\noindent where the spectral moments $\sigma^2_j(t,R)$ are
\bea \label{sigmaj}
  \sigma^2_j(t,R) &\equiv& \int_0^{\infty} \frac{\rmd k}{k} k^j \Delta^2_{\delta}(k,t) T(k)^2 W(k R)^2 \el
  &=& \frac{4}{9} \frac{1}{(\afrw \Hfrw)^4} \int_0^{\infty} \frac{\rmd k}{k} k^{j+4} \Delta^2_{\phi}(k) T(k)^2 W(k R)^2 \el
  &=& \frac{4}{9} \frac{ A^2}{(\afrw \Hfrw)^4} \int_0^{\infty} \frac{\rmd k}{k} k^{j+4} T(k)^2 W(k R)^2 \ ,
\eea

\noindent where $\Delta^2_{\delta}$ is the primordial power spectrum of density
perturbations, $\Delta^2_{\phi}$ is the primordial power spectrum of metric
perturbations, taken to be scale-invariant with amplitude
$A^2=(3\times 10^{-5})^2$,
$T(k)$ is the transfer function and $W(k R)$ is the window function,
taken to be Gaussian, $W(k R)=e^{-\ha k^2 R^2}$.
(When going from metric perturbations to density perturbations,
we have neglected the long-wavelength modes, as we are interested
in sub-horizon perturbations.)
The linear density field is taken to evolve in an
Einstein-de Sitter universe. We are neglecting the 'backreaction of
backreaction', the effect of the change of the average
evolution on the perturbations. One would need to
take this into account to see if there are, for example,
oscillations in the expansion rate as suggested in
\cite{Rasanen:2006a, Rasanen:2006b}.
For work on perturbations in a backreaction context, see
\cite{Sicka, Rasanen:2006b, Tatekawa:2001, Ehlers:1996, Takada:1999, Taruya:1999}.

To go from the peak number density to the fraction
of mass (or initial volume), we need to specify the mass associated
with each peak. We multiply the number density \re{n}
by the volume under the Gaussian window function, $(2\pi)^{3/2} R^3$,
and introduce a constant $N$ to account for the fact that the
total mass is not correctly normalised. The fraction of mass
in peaks of height $\nu=\delta/\sigma_0(t,R)$ is then
\bea \label{f}
  f(\nu,R) = N e^{-\frac{1}{2} \nu^2} \frac{1}{\sqrt{2\pi} (R_*/R)^3} G(\nu,\gamma) \ .
\eea

\noindent The total mass defined this way is
not constant with $R$. We fix the normalisation constant $N$
by demanding that asymptotically all mass resides in peaks
or troughs, $\int_{-\infty}^{\infty} \rmd\nu f(\nu,R)\rightarrow 1$
as $R\rightarrow\infty$. This gives $N\approx1.97$.
In this treatment, all peaks (and troughs) contain the same amount
of mass, regardless of their height (or depth).
A normalisation factor which depends on $\nu$ would probably be
more appropriate, but we want to keep the treatment simple.
The fraction of volume which is not in peaks or troughs is
taken to expand like the Einstein-de Sitter universe, $\Hfrw t=2/3$.
For discussion of mass assignment for peaks and troughs,
see \cite{Sheth, Bond:1991,Bond:1996}.

One factor which we have not taken into account is
the peak-in-a-peak problem.
The distribution function of peaks \re{f} does not take into
account that lower peaks may be submerged in higher ones.
There is an equivalent trough-in-a-trough problem, and perhaps most
importantly, the trough-in-a-peak problem, as some
underdense regions are extinguished by larger overdense
regions. There is no corresponding peak-in-a-trough problem,
and this asymmetry transfers mass from the
underdense to the overdense regions \cite{Sheth}.
Our treatment does not include this effect, and we have
equal mass in the underdense and overdense regions, since
the density field is Gaussian.

\para{Smoothing and time evolution.}

We determine the smoothing length $R$ by fixing $\sigma_0^2(t,R)$
to a given value at all times. We take this value to be unity,
$\sigma_0^2(t,R)=1$, so we have $\nu=\delta$.
Since the linear density contrast evolves in time,
$\sigma_0^2(t,R)\propto \afrw^2$, the smoothing length $R(t)$
will also evolve, growing with time. This in turn
translates into evolution of the distribution function $f(\d,R(t))$.
The smoothing scale $R(t)$ can be regarded as the typical size
of structures which are forming at time $t$.

All time evolution is taken into account in this statistical manner
in the distribution function $f(\d,t)$, since the individual regions
are by definition fixed at a given state of expansion or collapse.
This averaging of regional values of the expansion rate is somewhat
different from the Buchert formalism, where the basic quantity is the
local volume element. While the density parameters defined in
\re{omegas} can be used to understand the state of the universe
at each moment, the evolution of the expansion rate is not
completely captured by the Buchert equations, since the indirect
treatment of the local evolution of mergers by smoothing goes
outside the dust approximation.

In physical terms, smoothing with a window function involves the
assumption that we need not consider the details of structures
at the smoothing scale. It is a simplification for the merger of
structures, both overdense and underdense, into larger entities.
In the present setting, this is part of our assumption
of replacing continuous evolving structures by disjoint
spherical regions. In a full description of structures,
there would be a well-defined density contrast and expansion rate
at each point. When we model the universe with regions
which are each associated with a regional average density contrast
and expansion rate, smoothing is introduced.

\paragraph{The transfer function.}

The statistics of structures in the peak model are determined
by the functions $\gamma$ and $R_*$ defined in \re{gammaRstar}.
With a fixed primordial power spectrum, the behaviour of these
functions is given by the transfer function.

We will consider two approximations for the CDM transfer function.
The BBKS transfer function \cite{Bardeen:1986} is
\bea \label{BBKS}
  T(k) &=& \frac{\ln(1+2.34 q)}{2.34 q\left[1+3.89 q+(16.1 q)^2+(5.46 q)^3+(6.71 q)^4\right]^{1/4}} \ ,
\eea

\noindent with $q=k e^{\fb}/(13.7\keq)$ (assuming three massless
neutrino species), where $\fb\equiv\Ob/\Om$ is the baryon
fraction \cite{Weinberg:2002}
and $\keq^{-1}\approx 13.7 \om^{-1}$ Mpc is the wavelength of
the perturbations that enter the horizon during the matter-radiation
equality. Here $\om\equiv\Omn h^2$.
The BBKS transfer function is a fitting formula
to numerical results, and the large $k$ limit has been also
analytically derived \cite{Weinberg:2002}.
We fix the baryon fraction at $\fb=0.2$.
The results do not significantly depend on $\fb$
directly, but the BBKS approximation of the numerical
transfer function calculated with CAMB becomes worse with
increasing $f_b$; for $\fb=0.2$, the error at large $k$ is 20--30\%.

\begin{figure}[t]
\centering
\scalebox{0.5}
{\includegraphics[angle=0, clip=true, trim=0cm 0cm 0cm 0cm, width=\textwidth]{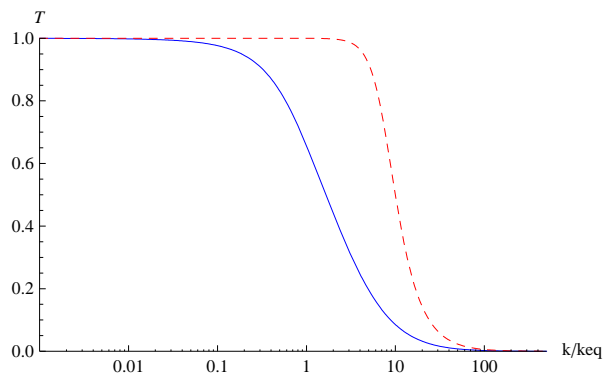}}
\caption{The BBKS (blue, solid) and BD (red, dashed) transfer functions as a function of $k/\keq$.}
\label{fig:transfer}
\end{figure}

For comparison, we will use the simple transfer function
introduced in \cite{Bonvin} by Bonvin and Durrer, which we
will call the BD transfer function,
\bea \label{BD}
  T(k)^2 = \frac{1}{ 1 + \b (k/\keq)^4 } \ ,
\eea

\noindent where $\b=3\times10^{-4}$.

The BD transfer function is a simple approximation to the
behaviour of the evolution of CDM perturbations captured in more
detail by \re{BBKS}. It is not quantitatively accurate, but
it will help to estimate the dependence of the results on
the transfer function.

Qualitatively, the BD transfer function has the same behaviour
as the BBKS transfer function. Perturbations with wavenumber
much larger than $\keq$ enter the horizon
deep in the radiation dominated era, so they are damped by $k^{-2}$,
while perturbations with much smaller wavenumber enter
in the matter-dominated era, so their amplitude is unsuppressed.
The  evolution in the slope of the BBKS transfer function is
shown in \fig{fig:transfer}.
Both the BBKS and the BD transfer function are missing the
cut-off at small scales due to free-streaming \cite{Boehm, SUSYCDM}.
However, we will see that the small-scale behaviour is not
important for our results.

\subsection{Preferred time in structure formation}

\para{The size of structures.}

Before evaluating the average expansion rate using the
peak model, we will briefly discuss the preferred time
that is involved in the formation of CDM structures from
a scale-invariant spectrum of primordial perturbations.
We would expect to see this timescale reflected in the
evolution of the expansion rate, regardless of the
statistical model used to describe structures.

Because of the change of the slope of the CDM transfer
function, the size of structures relative to the Hubble size
saturates around 10--100 billion years \cite{Rasanen:2006a, Rasanen:2006b}.
This can be seen as follows. Typical structures forming
at time $t$ have size $R(t)$ defined by $\sigma_0^2(t,R)=1$.
Inverting \re{sigmaj} to solve for $R$ and dividing by the 
Hubble radius, we get the relative size $R/(a H)^{-1}$
as a function of time.

The timescale is determined by the time of matter-radiation
equality $\teq\approx 1000 \om^{-2}$ years. This value
assumes three massless neutrino species with abundances
determined by thermal equilibrium, but
it is independent of late-time cosmology, as long as the energy
density of matter evolves like $a^{-3}$ and the energy density of
radiation evolves like $a^{-4}$.
In a clumpy space, the first follows from the
conservation of mass \re{cons}. There is no such conserved
quantity for radiation, so in general the energy density of
radiation does not necessarily evolve like $a^{-4}$ \cite{Buchert:2001}.
However, if the number density of radiation quanta is conserved
and their change in energy (i.e. redshift) is, at least on average,
related to the scale factor by $1+z=a^{-1}$,
the radiation energy density will be proportional $a^{-4}$.
This relates the studies of the passage of light to the study
of the Buchert equations in the case of non-zero pressure \cite{Buchert:2001}.

We adopt the value $\om=0.1$ for all plots,
giving $\teq\approx 10^5$ years. This value of $\om$ could
be reasonably moved up or down by a factor of 2,
so the timescale in the plots could be shifted
by a factor of 4 in either direction.
Model-independent estimates of $\Omn$ can be summarised
as $0.15\gtrsim\Omn\gtrsim0.35$ \cite{Peebles:2004}.
Regarding the Hubble parameter, the value from SNe Ia
observed with the Hubble Space Telescope has been quoted
as $h=0.72\pm0.08$ \cite{Freedman:2000}
or $h=0.62\pm0.05$ \cite{Sandage:2006}, depending
on the treatment of Cepheids (see section \ref{sec:obs}).
For comparison, the model-dependent value determined
from fitting the \LCDM model to the WMAP3 data
is $\om=0.127^{+0.007}_{-0.009}$ \cite{Spergel:2006}.

At early times, the $k^{-2}$ damping due to radiation domination
cancels the $k^2$ enhancement of the density perturbations
relative to the scale-invariant metric perturbations. Therefore the
spectrum of density perturbations depends only weakly on the scale,
so structures on different scales collapse nearly at the
same time. The initial structures are small and grow rapidly.
Beyond $\keq$, the amplitude is not suppressed, so after
the wavenumber of collapsing perturbations reaches $\keq$,
there is no scale in the system anymore, and
the relative size of the structures is constant.
The asymptotic size is roughly $\sqrt{A}\approx5\times10^{-3}$.

\begin{figure}
\hfill
\begin{minipage}[ht]{7.5cm} 
\scalebox{1.0}{\includegraphics[angle=0, clip=true, trim=0cm 0cm 0cm 0cm, width=\textwidth]{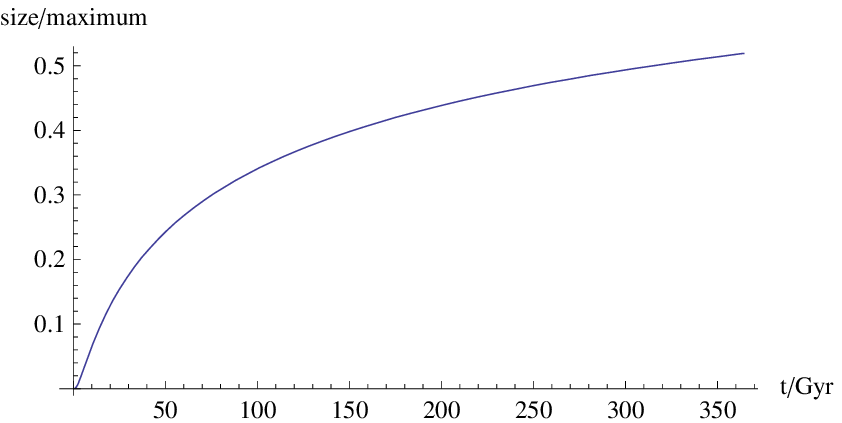}}
\begin{center} {\bf (a)} \end{center}
\end{minipage}
\hfill
\begin{minipage}[ht]{7.5cm}
\scalebox{1.0}{\includegraphics[angle=0, clip=true, trim=0cm 0cm 0cm 0cm, width=\textwidth]{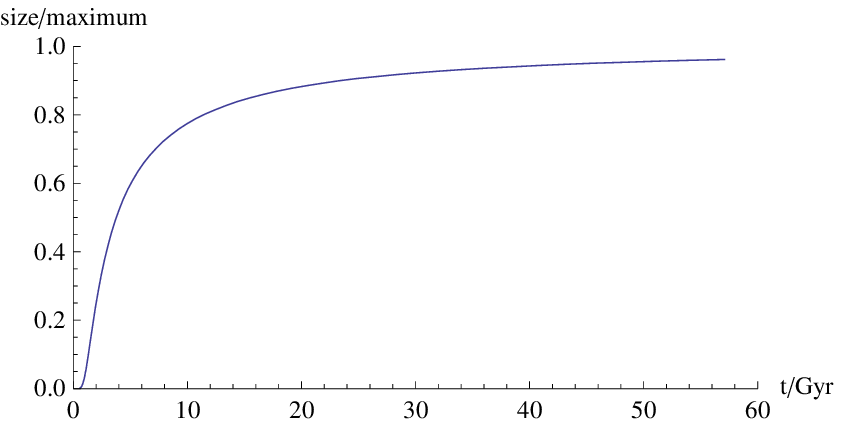}}
\begin{center} {\bf (b)} \end{center}
\end{minipage}
\hfill
\caption{The size of structures $R/(a H)^{-1}$  relative
to the asymptotic size as a function of time, in billions of years, for
(a) the BBKS transfer function and (b) the BD transfer function.}
\label{fig:size}
\end{figure}

We show the value of $R/(a H)^{-1}$ relative to its asymptotic
value in \fig{fig:size} for the BBKS and BD transfer functions.
The timescale of the evolution of the size is
sensitive to the behaviour of the transfer function
at small wavenumbers. For the BBKS transfer
function, the relative size enters the saturation regime
at some tens of billions of years.
For the BD transfer function, this happens at a few billion
years or so. In both cases, the evolution in
the size practically saturates well before 
the perturbations with wavenumber $\keq$ become non-linear,
which happens around 3500 billion years for the BBKS transfer
function and 2000 billion years for the BD transfer function.

The rough agreement of the era when structures reach their
maximum size with the time when acceleration has been observed
(about 10 billion years) is encouraging from the point
of view of the coincidence problem. A rough way to understand
the preferred time in terms of the time of matter-radiation
equality and the amplitude of the primordial perturbations
is as follows. The era when the structures
with wavenumber $\keq$ form is determined by the condition
$\sigma_0^2(t,\keq^{-1})=1$. Approximating the window function
$W(k R)$ with the step function, the transfer
function with unity for $k\leq\keq$ and recalling that
$\keq\equiv(a H)^{-1}_{\mathrm{eq}}$, we get from \re{sigmaj}
the time $t=(3/A)^{3/2}\teq\approx3000$ billion years.
The approach to the asymptotic value of the transfer function
is slow, and the saturation regime starts a couple
of order of magnitude earlier than this.
(It is apparent in the BD transfer function \re{BD} that
the turnover scale is $\approx10\keq$ rather than $\keq$.
Naively using $10\keq$ in the previous argument would bring
down the time to 3 billion years, since $t\propto (\afrw \Hfrw)^{-3}$.)
Exactly because the size changes slowly, it is
somewhat arbitrary what one calls the preferred time.
Around 10--100 billion years the relative size
of structures changes from growing rapidly to being almost
constant, but one cannot be more precise than that.

We will now get back to our model with the peak statistics,
where the expansion rate is determined as a function of time
without ambiguity, and see how it indeed changes around the saturation era.

\subsection{The expansion rate}

\para{Increasing $Ht$.}

The expansion rate is given by
\bea \label{Ht}
  \!\!\!\!\!\!\!\!\!\!\!\!\!\!\!\!\!\!\!\!\!\!\!\!\!\!\!\!\!\!\!\! H t = \int_{-\infty}^{\infty} \rmd\delta\, v_\d(t) H_\d t \el
   \!\!\!\!\!\!\!\!\!\!\!\!\!\!\!\!\!\!\!\!\!\!\!\!\!\!\!\!\!\!\!\!  = \frac{ \int_{-\infty}^0 \rmd\delta^{-}\, s_{\d^{-}} f(\delta^{-},R) H_{\d^{-}}t + \int_0^{\infty} \rmd\delta^{+}\, s_{\d^{+}} f(\delta^{+},R) H_{\d^{+}}t + \frac{2}{3} (1-\int_{-\infty}^{\infty} \rmd\delta\, f(\delta,R)) }{ \int_{-\infty}^0 \rmd\delta^{-}\, s_{\d^{-}} f(\delta^{-},R) + \int_0^{\infty} \rmd\delta^{+}\,  s_{\d^{+}} f(\delta^{+},R) + (1-\int_{-\infty}^{\infty} \rmd\delta\, f(\delta,R)) } \ ,
\eea

\noindent where $s_{\d^\pm}$ and $H_{\d^\pm} t$ are
given in \re{SCM} and \re{SEM},
and $f$ is given in \re{f}. The expansion rate is completely determined,
there are no free parameters to adjust (unless one counts the baryon
fraction in the BBKS transfer function). The average spatial curvature
$\av{\sR}$ and the backreaction variable $\sQ$ given in \re{Qpeak} are
calculated the same way.

\begin{figure}
\hfill
\begin{minipage}[h]{7.5cm} 
\scalebox{1.0}{\includegraphics[angle=0, clip=true, trim=0cm 0cm 0cm 0cm, width=\textwidth]{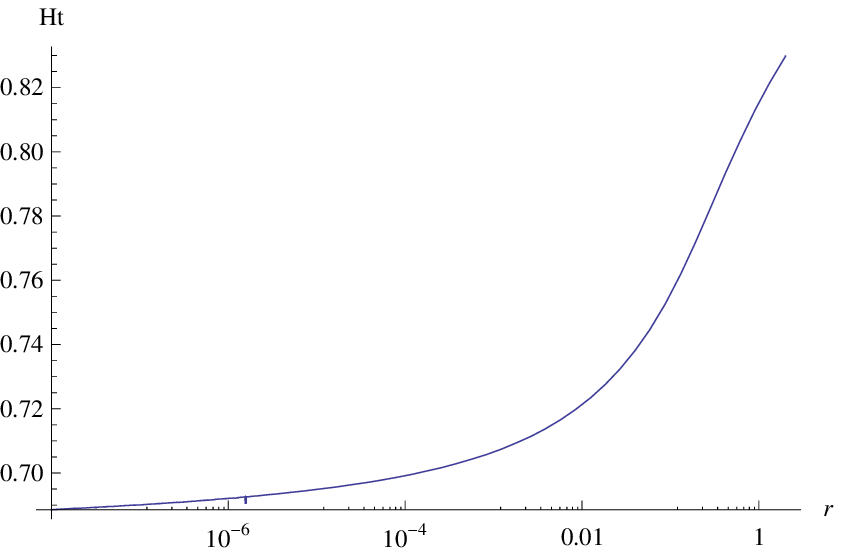}}
\begin{center} {\bf (a)} \end{center}
\end{minipage}
\hfill
\begin{minipage}[h]{7.5cm}
\scalebox{1.0}{\includegraphics[angle=0, clip=true, trim=0cm 0cm 0cm 0cm, width=\textwidth]{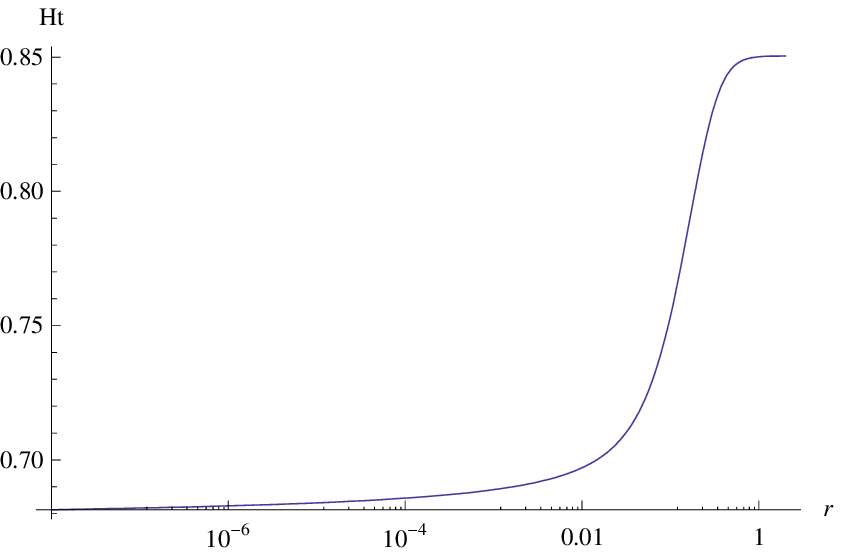}}
\begin{center} {\bf (b)} \end{center}
\end{minipage}
\hfill
\caption{The expansion rate $Ht$ as a function of $r=\keq R$ for
(a) the BBKS transfer function and (b) the BD transfer function.}
\label{fig:Htr}
\end{figure}

We first show $H t$ as a function of the size of structures
relative to the equality scale, $r\equiv \keq R$, in \fig{fig:Htr}.
Today $\sigma_0^2(t,R)$ is unity on the scale of 8 \mpc{} or slightly
below, so given $\keq^{-1}\approx13.7 \om^{-1}$ Mpc, we have
$r\approx\om\approx$ 0.05--0.2, placing the present time
in the transition region.
The transition era in the expansion rate is clear and the change
looks rapid, particularly for the BD transfer function. However,
because the growth of $r$ as a function of time slows down
as time goes on, the evolution is less steep as a function of $t$.

\begin{figure}
\hfill
\begin{minipage}[t]{7.5cm} 
\scalebox{1.0}{\includegraphics[angle=0, clip=true, trim=0cm 0cm 0cm 0cm, width=\textwidth]{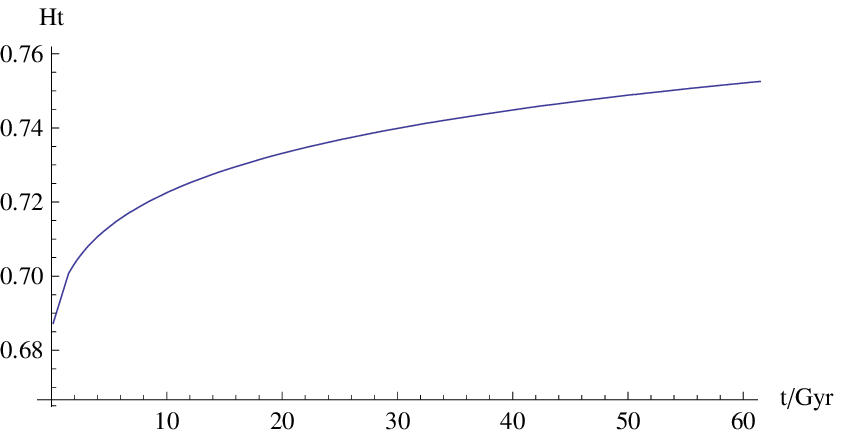}}
\begin{center} {\bf (a)} \end{center}
\end{minipage}
\hfill
\begin{minipage}[t]{7.5cm}
\scalebox{1.0}{\includegraphics[angle=0, clip=true, trim=0cm 0cm 0cm 0cm, width=\textwidth]{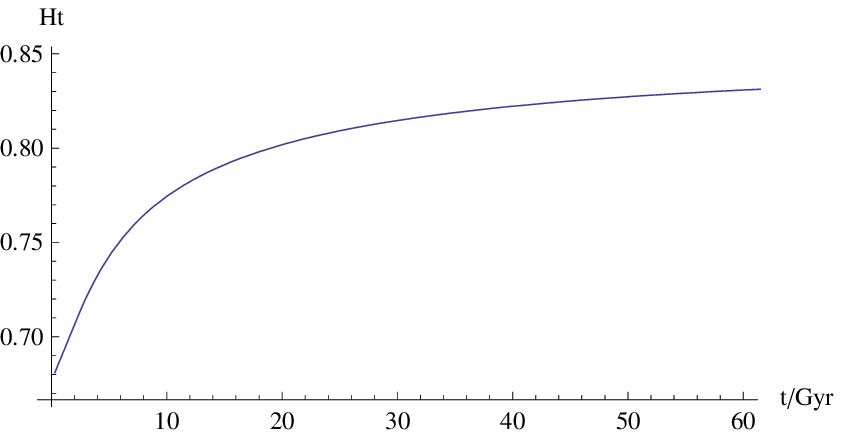}}
\begin{center} {\bf (b)} \end{center}
\end{minipage}
\hfill
\caption{The expansion rate $Ht$ as a function of time for
(a) the BBKS transfer function and (b) the BD transfer function.}
\label{fig:Ht}
\end{figure}

In \fig{fig:Ht} we show $Ht$ as a function of time.
The behaviour is qualitatively similar for the
BBKS and BD transfer functions. At early times,
$Ht$ is close to the FRW value $2/3$, slightly
higher because there are structures present.
With the onset of saturation (i.e. the turnover
in the transfer function), $Ht$ rises to a value
somewhat less than unity.

The physical reason for this evolution is that at early times
the volume occupied by the structures is small, so their impact
is small. As the volume occupied by structures grows
(along with the density contrast of typical structures),
the expansion rate becomes dominated by voids, since
their volume is large. This picture is in qualitative agreement
with excursion set studies of voids \cite{Sheth, vdW:2007}.
If all volume was in voids that were completely empty,
we would have $Ht=1$. Because the voids are not completely
empty, and because there are overdense
regions, the expansion rate asymptotes to a somewhat smaller value.
This evolution could be expected on general grounds,
as discussed in \cite{Rasanen:2006b}, but it is not
trivial that the timescale comes out correctly, in agreement
with the argument related to the size
of structures presented in the previous section.

As the Buchert equations \re{Ray}--\re{cons} show,
the effect of perturbations on the average expansion rate
does not necessarily become large when they first become non-linear.
The criteria for the breakdown of the perturbed FRW equations
as a description of the local evolution and as a description
of the average evolution are different. The latter breaks down
only when non-linear density perturbations occupy a sizeable
fraction of space\footnote{The time when backreaction
becomes important was argued in \cite{Notari:2005}
to be around today, by estimating when the perturbative corrections
to the average expansion rate become of order one. However, the
analysis does not take into account that the average
decomposes into a product of two-point functions, and that 
spatial derivatives have to occur in pairs. The correct order
of magnitude for the general term after eq. (8) in \cite{Notari:2005}
is $10^{-5} (a/a_0) \av{\delta^2}^{(n-2)/2}$. This does not select
out the present day.}.

\begin{figure}
\hfill
\begin{minipage}[h]{7.5cm} 
\scalebox{1.0}{\includegraphics[angle=0, clip=true, trim=0cm 0cm 0cm 0cm, width=\textwidth]{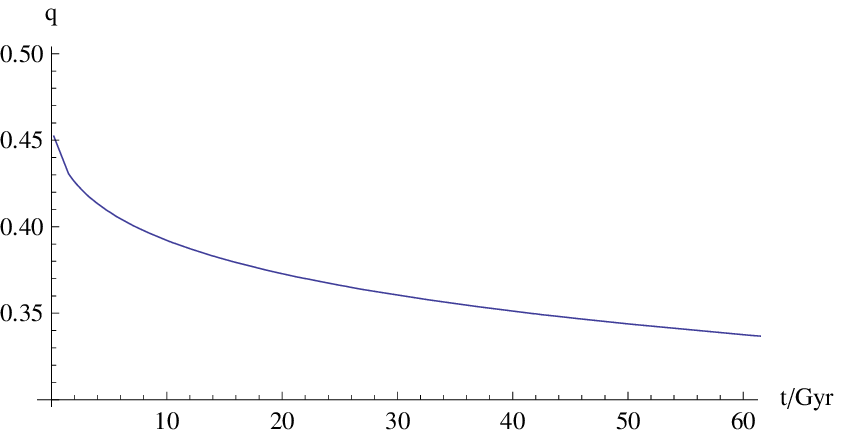}}
\begin{center} {\bf (a)} \end{center}
\end{minipage}
\hfill
\begin{minipage}[h]{7.5cm}
\scalebox{1.0}{\includegraphics[angle=0, clip=true, trim=0cm 0cm 0cm 0cm, width=\textwidth]{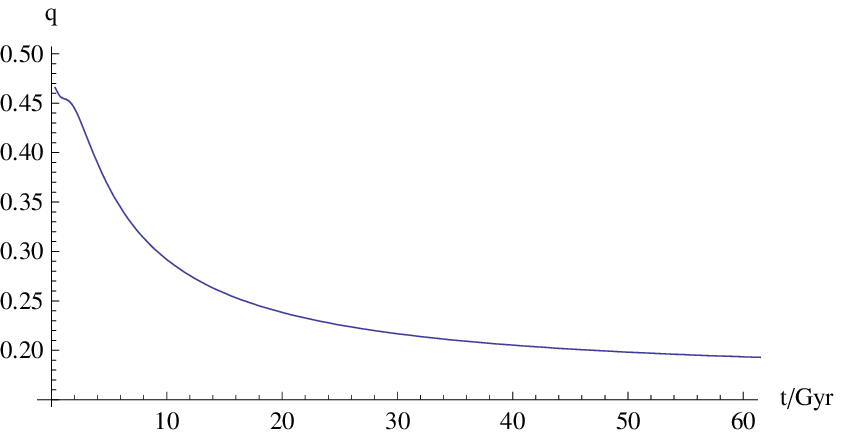}}
\begin{center} {\bf (b)} \end{center}
\end{minipage}
\hfill
\caption{The deceleration parameter $q$ as a function of time for
(a) the BBKS transfer function and (b) the BD transfer function.}
\label{fig:q}
\end{figure}

\para{No acceleration.}

In \fig{fig:q} we show the deceleration parameter
$q=-\addot/(a H^2)$ as a function of time.
Even though $Ht$ rises, there is no acceleration.
The reason is that only the underdense regions are
important, the overdense regions play almost no role. As a result,
$H t$ does not slow down before rising.
It is likely that in order to have acceleration, the 
overdense regions should first slow down the expansion rate, so that
the relative difference is larger, as in the toy model
discussed in \cite{Rasanen:2006a, Rasanen:2006b}.
In FRW models, there is a clear qualitative difference between
acceleration and deceleration, related to the dominant energy
condition.
Here the situation is different: acceleration is just
a quantitative question of the steepness of the $Ht$ curve
and the size of the density parameter $\OQ$, and there is
no principle involved.

\begin{figure}
\hfill
\begin{minipage}[h]{7.5cm} 
\scalebox{1.0}{\includegraphics[angle=0, clip=true, trim=0cm 0cm 0cm 0cm, width=\textwidth]{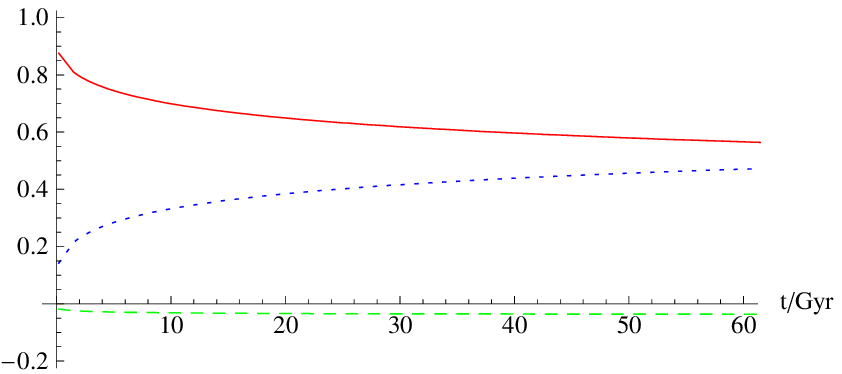}}
\begin{center} {\bf (a)} \end{center}
\end{minipage}
\hfill
\begin{minipage}[h]{7.5cm}
\scalebox{1.0}{\includegraphics[angle=0, clip=true, trim=0cm 0cm 0cm 0cm, width=\textwidth]{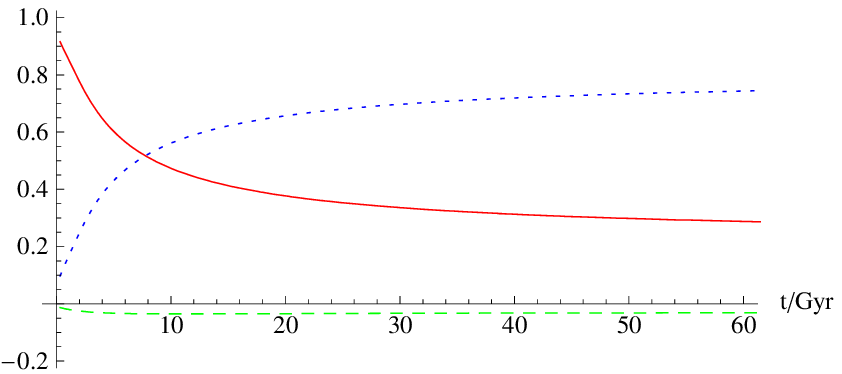}}
\begin{center} {\bf (b)} \end{center}
\end{minipage}
\hfill
\caption{The density parameters $\Om$ (red, solid), $\OR$ (blue, dotted)
and $\OQ$ (green, dashed) as a function of time for
(a) the BBKS transfer function and (b) the BD transfer function.}
\label{fig:omegas}
\end{figure}

In \fig{fig:omegas} we plot the density parameters $\Om$, $\OR$
and $\OQ$ defined in \re{omegas}.
For both transfer functions, the backreaction density
parameter $\OQ$ is small, below $0.04$ in absolute value.
(The backreaction contribution peaks in the transition era
when $Ht$ rises rapidly, but the change is too small to
appreciate in the plots.)
As noted in \cite{Sicka, Rasanen:2005},
the system can evolve far from the initial near-FRW
behaviour, even though $\OQ$ is small at each moment.
The backreaction density parameter $\OQ$ is a measure of the
distance of the model from the FRW case.
The smallness of $\OQ$ means that at each moment,
the system is near some FRW model and
evolves slowly between different near-FRW models.
(Note that the asymptotic value of $\OQ$ is non-zero, so
unlike an open FRW universe, the model does not become
emptier without limit: structures are present,
even when the volume is dominated by voids.)
In order for the behaviour at some instant to be very different
from a dust FRW model, $|\OQ|$ has to be large.
A backreaction variable $|\OQ|\gtrsim0.2$ is
needed to obtain acceleration of the observed magnitude \cite{Rasanen:2006b}.

While the timescale of the change in the expansion rate
comes out roughly in agreement with observations, the effect of
clumpiness is too small to give acceleration.
Let us now discuss the assumptions involved in the model,
and how to improve the treatment.

\subsection{Improving the model} \label{sec:impro}

\para{The spherical collapse/expansion model.}

Our model involves two parts:
structures are treated as an ensemble of regions whose
distribution is given by the peak model, 
and the evolution of the individual regions is taken to
follow the spherical collapse/expansion model.

Let us discuss the collapse/expansion model first. While the
spherical collapse model works surprisingly well for
the statistics of structures, it does not give an
accurate description of the collapsing phase.
In particular, there is no treatment of the stabilisation
which ends the collapse.
An effective treatment of the terms responsible
for stabilisation was given in \cite{Engineer:1998},
and a improved model which covers the whole evolution from
the linear regime to the stabilised phase was presented in
\cite{Shaw:2007}.
One can also generalise into ellipsoidal collapse, and take shear and
tidal effects into account \cite{Hoffman:1986, Bond:1996, DelPopolo:2001}.
The spherical collapse model in a backreaction
context has been discussed in \cite{Taruya:1999, Kerscher:2000}.

In general, a given value of the linear density contrast does not
correspond to a fixed value of the expansion rate, unlike in the
spherical collapse model.
In terms of the Buchert equations \re{Ray}--\re{cons},
the backreaction variable $\sQ$ is non-zero,
and the equations do not have a unique solution.
However, one could generalise the present treatment to
cover more realistic structures  by adding one layer
of ensemble.
One could integrate over a distribution of initial conditions
to obtain the mean expansion rate corresponding to structures
of a given density contrast, using the Buchert equations, as
presented in \cite{Kerscher:2000, Sicka}.
While the approximation that the expansion rate depends
only on the density contrast (used in \cite{Engineer:1998, Shaw:2007})
is probably not valid for a single structure, it is true by
construction for the expansion rate averaged over
different structures.

There is no stabilisation problem with the underdense
equivalent of the spherical collapse model, and the evolution
smoothly approaches the empty case $H_{\d^-} t=1$. However, the
model breaks down if shells from inner regions of the void
(which have a smaller density and therefore expand faster)
catch up with outer shells, even though this is not apparent
in the average description. When this happens
depends on the initial density profile. For a top-hat
density profile, shell-crossing occurs at linear density contrast
${\d^-}\approx-2.8$, corresponding to $H_{\d^-} t=8/9$ \cite{Sheth}.
In real structures, the shell-crossing singularity
corresponds to the formation of a dense wall surrounding
the void. After wall formation, it might be more
realistic to take the expansion rate to be that of an
outward-moving shell, $H_{\d^-} t\approx0.8$ \cite{voids, Bertschinger:1985}.
Like in the case of overdense regions, 
the assumption of spherical symmetry may be questionable.
While isolated voids grow more spherical \cite{Bertschinger:1985, Icke:1984},
the shapes of real voids are affected by surrounding structures.
Voids in simulations have complicated shapes
\cite{Shandarin, Platen:2007}, though the overall structure
resembles a foam of spherical voids \cite{Sheth}.
The shapes of underdense structures could be taken statistically
into account with an ensemble of initial conditions, as with
the overdense regions.

However, refinements to the collapse/expansion model are secondary
to the issue of distribution of the overdense and underdense regions.
For example, redoing the calculation above with the improved
collapse model presented in \cite{Shaw:2007} makes practically
no difference to the results. The reason is that the
volume in the overdense regions is so small that their impact
is negligible. In order to have slowdown, and after that
acceleration, the fraction of the initial mass (or,
equivalently, initial volume) which goes into the overdense
structures should be larger.

\para{The peak model of structures.}

We have already mentioned that the distribution function \re{f}
does not take into account the peak-in-a-peak problem,
trough-in-a-trough problem or trough-in-a-peak problem.
In particular, the last mentioned involves the extinction of
voids by collapsing structures, which leads
to the transfer of mass from underdense to overdense
regions, increasing the volume fraction of the latter.
In the excursion set formalism, the
extinction of voids by overdense structures is crucial 
for getting a single void size to dominate at each epoch
\cite{Sheth, vdW:2007}, as observed \cite{voidobs}.

The peak model is based on an isolated view
of structure formation, in which matter concentrations
remain stationary, with mass accreting onto
(or flowing away from) the fixed extremal points of the
initial density field. Shear and tides are neglected.
However, such effects can be important, and the sites of
structure formation do not necessarily coincide with the
initial density peaks (even underdense regions can collapse)
\cite{Hoffman:1986, Bond:1996, Kerscher:2000, Platen:2007}.

The mass assignment of the peaks and troughs is likely to
be important, and one should consider a realistic
mass function instead of having all peaks and troughs
contain the same amount of mass.
Certainly, the treatment of merging and assignment of mass
with the simple Gaussian smoothing is somewhat arbitrary,
and the mass assignment should reflect the
dynamics of structure formation.
A realistic treatment would be expected to break
the symmetry in mass between peaks and troughs.

We have said that our model does not have new free parameters,
and this is true in the sense that we have not adjusted anything
to get a certain result. However, the smoothing threshold,
which we set at $\sigma_0(t,R)=1$, could have been chosen
to have some other value.
The peak distribution \re{f} depends on $\delta/\sigma_0(t,R)$,
and increasing $\sigma_0(t,R)$ enhances the effect of structure.
However, changing $\sigma_0(t,R)$ around unity does not make
much difference.
Setting $\sigma_0(t,R)=2$ increases $Ht$ by less than 10\%,
and the choice $\sigma_0(t,R)=0.5$ brings $Ht$ down by
around 10\%.
The rate of change, i.e. the steepness of the $Ht$ curve,
increases with $\sigma_0(t,R)$, and the deceleration parameter
$q$ changes by tens of percent when setting $\sigma_0(t,R)$
to 2 or 0.5. The time when the expansion rate changes significantly
remains about the same, and the qualitative behaviour is unchanged.

Finally, the results depend strongly on the transfer function,
and the BBKS approximation may not be sufficient near the important
bend in the spectrum; this is easy to correct in a numerical
treatment. The results are not sensitive to the transfer function
and power spectrum at small scales, where they are poorly known.
The reason is that at early times when small structures form,
their effect on the expansion rate is small.
On the other hand, rapid changes in the power spectrum
at large scales could have a large impact.
Changing the primordial spectrum from scale-invariant
to a power-law with a spectral index of $n=0.8$ or $n=1.2$
(with a pivot scale of $0.002$ Mpc$^{-1}$) changes $Ht$ by
only a few percent. More drastic changes in the power spectrum,
such as a sudden transition, would be needed for
a significant effect.

A more realistic treatment of the effects discussed above
will make it possible to say with more confidence whether
overdense regions are prominent enough to give
acceleration. If structures indeed lead to to
accelerated expansion, the Buchert equations
\re{Ray}--\re{cons} dictate certain overall features
that the universe must have, independent of the details
of the structures. It is worth considering the compatibility
of these general features with observations, as they might
in principle rule out backreaction as an explanation for the
observations. (See \cite{Rasanen:2006b} for further comparison
to observations.)

\section{Discussion}

\subsection{Observational issues} \label{sec:obs}

\para{The age of the universe.}

Observationally, the age of the universe is not known
with precision in a model-independent manner.
There is an important constraint from the ages of
globular clusters, which give the lower limit $t_0\geq11.2$ Gyr at 95\% C.L.
and a best-fit age of $t_0=13.4$ Gyr \cite{Krauss:2003}.
These results do not depend on the details of late-time cosmology,
and should be valid as long as the expansion at redshifts $z>6$
is close to the Einstein-de Sitter case.
For the Hubble parameter, the most accurate model-independent
determination is from SNe Ia observed with the Hubble Space
Telescope \cite{Jackson:2007}.
The usually quoted value is $h=0.72\pm0.08$ \cite{Freedman:2000}
(in close agreement with $h=0.73\pm0.06$ found in \cite{Riess:2005}),
while recent work with a different treatment of Cepheid metallicity
gives $h=0.62\pm0.05$  \cite{Sandage:2006} (all 1$\sigma$ limits).
These estimates give a mean value of $H_0t_0\approx0.99$
or $H_0t_0\approx0.85$, respectively, using $t_0=13.4$ Gyr.
Using the lower limit for $t_0$ and the mean value for
$h$ gives $H_0t_0\gtrsim0.83$ or $H_0t_0\gtrsim0.71$, respectively.
The treatment of Cepheids does not appear to be settled
\cite{cepheid, Jackson:2007}, but in any case,
$H_0 t_0$ values in the lower range are not ruled out.
In principle, $H_0t_0$ can provide an important
constraint on backreaction models.
In particular, a definitive measurement of $Ht>1$
would imply that exotic matter is needed,
because in a dust universe $Ht\leq1$, regardless
of the structures present (assuming that vorticity
can be neglected) \cite{Rasanen:2005}.

The values of $H t$ today in \fig{fig:Ht} are
rather low for the BBKS transfer function,
while the BD transfer function gives results more
in agreement with the observations.
The difference is mainly due to the fact that the
change in the fraction of mass which is in structures
is slower in the BBKS case. For the BBKS transfer function,
only 30\% of the mass is in structures at 15 Gyr, compared
with 54\% for the BD transfer function. In both cases, initially (i.e. in
the limit $t\rightarrow0$) the fraction is 10\%.

\para{The CMB peaks.}

As the exact expression \re{RfirstQ} shows and
\fig{fig:omegas} demonstrates, the average
spatial curvature is generically large in a dust
universe with significant clumpiness.
This raises the question of
compatibility of the backreaction explanation
for the acceleration with CMB observations.
However, the fact that the CMB data is in good agreement
with FRW models that have no spatial curvature does not
imply that the CMB rules out models with spatial curvature.
Conclusions drawn about spatial curvature from the
CMB are model- and prior-dependent.

In physical terms, the CMB peak structure is mostly determined
by the primordial spectrum of perturbations, by $\Om h^2$ and $\Ob h^2$,
and by the shift parameter, which is a measure of the distance to
the last scattering surface \cite{Efstathiou:1998, Mukhanov:2003}.
The CMB anisotropies will be the mostly the same in models with
identical values of these parameters, apart from the low
multipoles (though see \cite{Elgaroy:2007, Corasaniti:2007}).

Even in the \LCDM model with primordial perturbations given by a
power law, the CMB alone does not constrain spatial curvature 
to be small \cite{Spergel:2006}, though combining the CMB
with a measurement of Hubble parameter does provide
a strong constraint.
In a FRW model with a time-varying equation of state,
at least $\Omega_K\approx0.2$ is compatible with
the correct CMB shift parameter (and fitting the SN Ia data
and the $A$-parameter from baryon acoustic oscillations) \cite{Ichikawa}.

In a backreaction model, there is no simple argument for
obtaining the position of the CMB peaks.
Both positively and negatively curved regions are present, and
there may be more cancellation in the effect on the passage
of light than would be expected based on a FRW model with
the equivalent amount of spatial curvature.
One has to redo the CMB analysis, considering the passage
of light in a universe with realistic structures, as
discussed in section \ref{sec:prop}.
It might even be that the change in the passage of light together
with the increase in $Ht$ could explain the observations without
acceleration, as discussed in \cite{Mattsson:2007b}.

In any case, the contribution of spatial curvature
is regionally not negligible in the real universe.
(The following discussion neglects effects due to breakdown
of the dust approximation.)
For example, consider an overdense dust structure.
For a shell that is turning around from expansion
to collapse, the left-hand side of \re{Hamloc} is zero.
Neglecting vorticity (in the spherically symmetric case,
it would be zero), the spatial curvature must
be positive, and equal to the
sum of the contributions of the energy density and shear,
$\sR=16\pi\GN\rho+2\sigma^2$.
For a stabilised structure, we have $\theta=\dot\theta=0$,
so solving from \re{Rayloc} and \re{Hamloc} we
obtain $\sR=12\pi\GN\rho$ and $\omega^2=2\pi\GN\rho+\sigma^2$.
Just as vorticity is needed to balance against the energy
density and shear to have zero acceleration, zero expansion
requires the spatial hypersurfaces to be positively curved.
The energy density of stabilised structures can
be much larger than the average energy density,
so the same is true of their spatial curvature.
(The growth in the relative
spatial curvature is balanced by the fact that the
volume occupied by stabilised regions shrinks by the
same factor that their energy density increases, so
the contribution to the average Hubble rate \re{Ham}
remains constant.)

Inside voids, the energy density is much smaller than the
background value, while the expansion rate is larger,
so the contribution of the spatial curvature is
again larger than that of energy density.
For example, consider a void on an Einstein-de Sitter
background. Neglecting shear and vorticity, we have
$|\sR_{\mathrm{void}}|>16\pi\GN |\delta|\av{\rho}$,
where $\delta$ is the real (not linear) density contrast of the void
(typically $\delta\approx-0.9$ for observed voids \cite{voidobs}).

In principle, observations of structures correlated with
the CMB at different redshifts could lead to useful constraints
on the evolution of the spatial curvature with redshift, as
with the ISW effect \cite{Ho:2008}, once the effect of
non-linear structures on the passage of light is properly understood.

\para{Variance.}

A variance $|\OQ|\gtrsim0.2$ is required to explain
the observed acceleration \cite{Rasanen:2006b}.
The directional variation of the expansion rate
was studied in \cite{McClure:2007} using SNe Ia
(following the treatment of \cite{Freedman:2000}
rather than \cite{Sandage:2006}).
Typical variation was found at the 10--20\% level,
the maximum difference in the expansion rates being over 50\%.
Of course, angular variation is different from volume
variance, and the systematics of SNe Ia are perhaps
not completely understood, as discussed in section \ref{sec:intro}.
For directional analysis of SNe Ia, see also \cite{Haugbolle:2006, Schwarz:2007}.

In \cite{Li:2007b}, the directional variation of the expansion
rate inferred from SNe Ia in \cite{Freedman:2000} was interpreted
as evidence for backreaction. However, the estimate 
of the variance in \cite{Li:2007b} is unreliable (even apart
from the applicability of linear perturbation theory).
The magnitude of the the relevant term
$B=\av{\pat_i(\pat_i\varphi\nabla^2\varphi)}-\av{\pat_i(\pat_j\varphi\pat_i\pat_j\varphi)}-\frac{2}{3}\av{\nabla^2\varphi}^2$
was estimated by replacing spatial derivatives with the inverse
of the averaging scale, and replacing the two powers of $\varphi$
by the primordial amplitude of the power spectrum. However,
spatial gradients are determined by the scale over which the
perturbations vary significantly (as determined by the power
spectrum and the transfer function), not by the averaging scale.
For example,
$\av{\delta^2}\propto\av{\nabla^2\varphi\nabla^2\varphi}$
is divergent for a scale-invariant spectrum (with no free-streaming
cut-off), regardless of how large the averaging scale is, because
there are perturbations on arbitrarily small scales.
Conversely, with a free-streaming cut-off, the quantity is finite
no matter how small the averaging scale is.

Note that the variance in the expansion rate evaluated from
Newtonian simulations \cite{Rauch:2005} is large enough
to provide acceleration, were it not balanced
by shear, as noted in \cite{Rasanen:2006b}.
As we discuss in  section \ref{sec:Newton} below,
such a cancellation is not in present in general relativity.
If the variance in simulations were negligible,
it would be less plausible that it is large in the real
relativistic case. However, the Newtonian variance is large,
so in order to have acceleration, the shear simply has to be
smaller than in the Newtonian case.

It is sometimes argued that the dust shear would be of
the order $\sigma^2\lesssim 10^{-10} H^2$ from the isotropy of
the CMB \cite{EGS} (see also \cite{moreEGS}).
Any real non-linear dust structure violates this bound, so
it is clear that the bound does not apply to the real universe.
(The calculation also gives a limit of
$\abs{\nabla\rho/\rho}\lesssim 10^{-5} H$ for the 
spatial scale of density variation, which is even more obviously violated.)
However, if the shear really was negligible, then
this would indicate that backreaction is important, as
observations and simulations suggest a significant
variance of the expansion rate \cite{Rasanen:2006b}.

\para{Coldness of the Hubble flow.}

We have noted that there is no evidence for a matter component
with negative pressure apart from the cosmological observations
of accelerated expansion. In contrast to this, it has been claimed
that the influence of vacuum energy can be seen in the local
dynamics within the local 10 Mpc or so from the
'coldness' of the local Hubble flow.

The coldness refers to two distinct phenomena\footnote{The
linearity of the nearby mean flow with distance is sometimes
mentioned as a third aspect.}.
First, the velocity dispersion in the local volume has been claimed
to be anomalously low. Some authors find the dispersion
within the nearest few Mpc to be $\approx$ 40 km/s \cite{lowsigma},
while others find a value of 88 km/s $\pm$ 20 km/s \cite{Maccio:2004},
or over 100 km/s \cite{Whiting:2002}.
The second aspect is that
the expansion rate measured within the nearest 10 Mpc
is said to agree quite well with the global Hubble rate.
Given that the matter distribution is locally quite clumpy,
and does not become homogeneous until around 100 \mpc,
this seems somewhat surprising.
It has been suggested that both observations are explained
by vacuum energy. It can 'cool' the local expansion rate, and if
vacuum energy dominates also the global dynamics, it is natural
that the local and global expansion rates agree.

The velocity dispersion argument was presented as evidence against
a high density matter-dominated FRW model before the SN Ia
observations supported accelerating expansion \cite{Suto}.
It has been argued that the observed low value
of the velocity dispersion could be due to the known
peculiarity of the local structures instead \cite{vdW},
or even that such a small dispersion is typical \cite{Leong:2003}.
However, it seems that the typical velocity dispersion
in simulations of the EdS model is indeed considerably higher,
300--700 km/s \cite{Governato:1996}, while \LCDM simulations
can reproduce the small velocity dispersion \cite{Klypin:2001}.
In simulations constrained to reproduce the large
scale structure of the local universe, the velocity dispersion
around regions similar to the Local Group of galaxies has been found
to be as cold in the open CDM model as in the \LCDM model,
given the same matter density \cite{Hoffman:2007b}.
This still leaves the question of how common such regions are.
However, the lower range of the values 150--300 km/s
found in unconstrained simulations of the open CDM model \cite{Governato:1996}
is not very dissonant with an observational value of over 100 km/s.
It does not seem unreasonable that the effect of spatial
curvature would be even stronger in a clumpy model,
where the density parameter of spatial curvature can be larger.

Unlike for the velocity dispersion, the probability of being located
in a region where the local expansion rate agrees with the global
value has not been studied quantitatively. At any rate, as
long as the measurement of the global and local Hubble parameters
is uncertain, it seems difficult to draw strong conclusions.
The notably different values $h=0.62$ and $h=0.72$ have both
been used as evidence for the vacuum energy origin of the
local expansion rate \cite{Sandage:2006, Chernin, localde}.

It has also been argued that the influence of vacuum
energy has been directly measured in the motions of
nearby galaxies \cite{Chernin, localde}. The idea is
that for a spherically symmetric system with a large
central mass, the repulsive gravity of vacuum energy
dominates beyond a certain radius, estimated as 1--2
Mpc for the Local Group. However, the assumption of
a spherically symmetric field generated by a point mass
used in the analysis does not hold very well, and the observations
are also consistent with domination by negative curvature \cite{Hoffman:2007b}.

\subsection{Non-Newtonian effects} \label{sec:Newton}

\para{Spatial curvature and Newtonian gravity.}

It is not clear how much the overdense regions slow down the
expansion rate, but from the model we have discussed, it seems
difficult to avoid the conclusion that underdense regions
cause $Ht$ to increase significantly.
The physical interpretation seems straightforward: the
fraction of space in faster expanding regions grows.
However, there is a subtlety involved.
The evolution we took for the individual regions
is purely Newtonian, and the peak statistics
do not involve general relativity. Had we
formulated the problem in Newtonian gravity instead
of general relativity, the answer would have been the same.
However, in Newtonian gravity, the backreaction variable $\sQ$
is a boundary term \cite{Buchert:1995}, so backreaction
vanishes for periodic boundary conditions. (In particular,
this is true for Newtonian simulations.)

While the universe presumably does not have periodic
boundary conditions at the visual horizon, the result
implies that backreaction in Newtonian gravity also vanishes
for a statistically homogeneous and isotropic system. This can
be seen in two ways. First, consider a volume which has periodic
boundary conditions on a scale much larger than the
homogeneity scale. By statistical homogeneity and isotropy,
the value of $\sQ$ evaluated within each homogeneity scale
sized box in the volume is the same as the overall value,
which is zero due to the boundary
conditions. Because the local physics is independent of the
boundary conditions on very large scales, this result should
hold even if the boundary conditions are not periodic\footnote{I
am grateful to Christof Wetterich for this argument.}.
A perhaps more physical argument, explained in \cite{Notari:2005},
is that a boundary term can be viewed as a flux going from one
region to another, and in a statistically homogeneous and isotropic
region, there should be an equal flux in and out.

In Newtonian gravity, the average evolution of a statistically
homogeneous and isotropic dust space does always follow the FRW
equations (i.e. the Buchert equations with $\sQ=0$).
However, this is not true in general relativity.
The Newtonian theory constraint that variance and shear
in $\sQ$ in \re{Q} cancel up to a boundary term is not
present in general relativity \cite{Buchert:1999}.
Viewed equivalently in terms of the spatial curvature
via the integrability condition \re{int}, this difference
is a reflection of the absence of spatial curvature in Newtonian gravity.

In Newtonian space and time, the geometry of the spatial hypersurfaces
is fixed, so the spatial curvature tensor and its trace, the
spatial curvature scalar $\sR$, do not exist \cite{Ellis:1971}. 
Therefore the Newtonian equations of motion do not include
an equation for the expansion rate such as the Hamiltonian
constraint \re{Hamloc} independent of the Raychaudhuri equation.
The analogue of the relativistic Hubble equation \re{Hamloc}
emerges only as the first integral of the Raychaudhuri equation.

In Newtonian gravity, the backreaction variable $\sQ$ is
given by a boundary term which is in general non-zero.
However, for an isolated system or one that is
statistically homogeneous and isotropic, we have $\sQ=0$.
Then the the first integral of the Newtonian equivalent of the
average Raychaudhuri equation \re{Ray} gives an equation like
the average Hamiltonian constraint \re{Ham}, but with a term
proportional to $a^{-2}$ in place of $\av{\sR}$, as \re{firstQ} shows.
The $a^{-2}$ term multiplied by $a^2$ can be interpreted
as the conserved energy of the Newtonian system.

There is no such conserved quantity in the relativistic case.
The average spatial curvature can evolve non-trivially,
unlike the total energy of an isolated Newtonian system.
The average spatial curvature is constrained by the
integrability condition \re{int} for the average
Hamiltonian constraint and the average Raychaudhuri equation,
instead of being proportional to $a^{-2}$.
(In the FRW case, the relativistic spatial curvature term and
the Newtonian energy term are mathematically identical,
only the physical interpretation is different.)

In our model, the evolution of overdense and underdense
regions is uncorrelated. We do not have a Newtonian constraint
on their behaviour, and the overall average spatial curvature
can evolve non-trivially, as in the general relativistic case.
A Newtonian calculation for a statistically homogeneous
and isotropic system would have to include the global constraint
that the underdense and overdense regions add up in such a manner
that the energy term (corresponding to the spatial curvature)
is proportional to $a^{-2}$.
(In a consistent treatment with a continuous distribution
of matter, instead of isolated regions, this would
follow automatically.)

There is a similar situation in the case of spherical symmetry.
As noted in section \ref{sec:setup}, the backreaction variable
$\sQ$ vanishes for spherical symmetry in Newtonian gravity
\cite{Sicka, Buchert:2000}, resulting in the spherical collapse/expansion
model. In general relativity, spherical symmetry does
not imply perfect cancellation between variance and shear,
and the average expansion of a spherically symmetric dust solution
can even accelerate \cite{Chuang:2005, Kai:2006, Paranjape:2006a}.

\para{The Newtonian limit of general relativity.}

In order for backreaction to be able to account for
the observed acceleration, non-Newtonian aspects of gravity
should be important already at the homogeneity scale of
around $100$ \mpc.
It is sometimes argued that general relativistic effects
in the present-day universe can only be important on
super-Hubble scales or near neutron stars and black holes.
However, this conclusion is based on
linearly perturbed FRW or Minkowski universes or the
Schwarzschild solution, and the general situation is not that simple.
(The subtleties of the Newtonian limit of near-FRW cosmological
models have been discussed in \cite{Ellis:1994, vanElst:1998};
see also \cite{Kofman:1995} regarding the magnetic part
of the Weyl tensor.)

The Einstein equation has ten components, with four constraints,
whereas in Newtonian gravity there is only one equation,
the Poisson equation. There are two aspects to the
additional equations. Some of them correspond to
degrees of freedom which do not exist in Newtonian
gravity (such as gravity waves and spatial curvature),
while others provide new constraints on the existing
degrees of freedom. 
The low-velocity, weak-field limit of general relativity
is not Newtonian gravity, but Newtonian gravity with
extra degrees of freedom and additional constraints.

General relativity reduces to Newtonian gravity in the limit
of infinite speed of light. However, this limit is not smooth.
Taking the speed of light to infinity removes some
equations completely and makes Newtonian gravity
qualitatively different from general relativity,
where the speed of light is finite.
Relativistic effects do not require velocities near
the speed of light, only a finite speed of light.

For example, there exist solutions of Newtonian gravity
which are not the limit of any general relativity solution,
due to additional constraints in the relativistic case. 
There are Newtonian expanding dust solutions
which have zero shear but non-zero vorticity.
However, in general relativity, non-zero vorticity
implies non-zero shear for expanding dust \cite{Ellis:1967, Ellis:1971}.
This does not require large velocities or strong gravitational fields;
see \cite{Senovilla:1997} for a clear comparison of the
general relativistic and Newtonian cases.

In the case of backreaction, it is the presence of new
degrees of freedom, namely spatial curvature, rather
than extra constraints which is important.
In perturbation theory around FRW universes, the leading
terms in the backreaction variable $\sQ$ are Newtonian,
so they are total derivatives and do not contribute in a
statistically homogeneous and isotropic
system \cite{Rasanen, Kolb:2004a, Notari:2005}. 
The non-Newtonian terms do not appear earlier than fourth
order in perturbation theory. Their numerical coefficients have
not been evaluated (it may be possible to do so using
third order perturbation theory, instead of having to calculate
to fourth order \cite{Li:2007a}), but their form is known, and
velocities near the speed of light are not required for them to be important.

\para{Simulations and backreaction.}

If non-Newtonian aspects are important already at the homogeneity
scale, one would expect them to show up also in quantities other
than the expansion rate.
We can ask whether there is any room for such effects,
given the comparison of simulations of structure formation
(which are completely Newtonian) with observations.
In fact, while simulations reproduce many features of
observations, there are some notable differences.

In \cite{SylosLabini:2006} it was found that the homogeneity scale
evaluated from simulations was only of the order 10 \mpc,
an order of magnitude smaller than the observed
70--100 \mpc \cite{Hogg:2004, Pietronero}.
(The box size of the simulations studied in \cite{SylosLabini:2006}
was 141 \mpc, so the correct homogeneity scale could not have been
reliably recovered. However, the homogeneity scale that was found
is much smaller than the box size, so it is not likely to be
a finite size artifact.)

According to \cite{Einasto},
the number of largest structures in the Millennium Simulation
is underproduced compared to observations, by a factor of 10
for the most luminous superclusters.
As for large underdense structures, the voids produced
in simulations are not as empty as those observed. This
`void phenomenon' has even been called a crisis of the
\LCDM model \cite{Peebles}.
(It has recently been argued that simulations
coupled with semianalytical galaxy formation in fact
agree with the void observations \cite{vonBendaBeckmann:2007}.)

These discrepancies may have resolutions which
have nothing to do with non-Newtonian physics or backreaction.
The statistics for the largest structures are probing the tail
of the distribution, which may be affected by non-linear
corrections to the evolution of the power spectrum
\cite{Crocce:2007} or transients from initial conditions \cite{Crocce:2006}.
The magnitude of such effects has been found at the 10--30\% level,
and the discrepancy is an order of magnitude, but
one should not be confident that there are no
unaccounted for effects in simulations, for example
related to discretisation \cite{discrete}.
The apparent emptiness of voids, in turn, might be due to
bias or subtleties in galaxy formation and the definition
of voids, rather than spatial curvature \cite{Ostriker:2003, Furlanetto:2005}.

Nevertheless, the point is that we do not have confirmation
that Newtonian gravity works well near the homogeneity
scale, since the Newtonian results differ in some important
respects from the observations.

\subsection{Conclusion}

\para{Summary.}

We have studied the effect of structure formation on
the expansion rate in a statistically homogeneous and
isotropic space.
We first reviewed studies of the propagation of light
in a space with non-linear structures, and discussed
some qualitative issues related to the average expansion rate.
We then calculated the average expansion rate in a model
where the number density of structures is given by the
peak model of structure formation with cold dark
matter, and the individual structures are described
with the spherical collapse model and its underdense equivalent.
In the calculation, there are no adjustable free parameters
in addition to those related to structure formation in a
spatially flat matter-dominated FRW universe.

We find that the expansion rate increases relative to the FRW value
at a time of tens of billions years, about the observed acceleration
era, possibly offering a solution to the coincidence problem.
The timescale has its origin in the change of slope
of the CDM transfer function around the matter-radiation
equality scale $\keq$. This leads to a preferred time around
$10^5\teq\approx 10^{10}$ years, when the volume occupied
by structures and their size relative to the visual horizon
saturate, and the impact of structures on the expansion rate becomes large.

However, while $H t$ increases in the model, it does not
rise sufficiently rapidly to correspond to acceleration.
The expansion rate increases because the
relative volume of the faster expanding regions rises,
as quantified by the Buchert equations. In order to have
acceleration, it is likely that the average expansion rate
would first have to be damped by slower expanding overdense
regions before it is increased by shedding
their contribution when voids become dominant.
In the present model, this does not happen, because
the volume occupied by the overdense regions is rather small.
This is partly due to the fact that we have equal amounts of mass
in overdense and underdense regions. Taking into account mass flow
to the overdense regions would increase their impact.

Improving the treatment of the statistics of the peaks
and the model used for individual structures would lead
to a more realistic estimate of the expansion rate.
The propagation of light in a universe with realistic,
evolving structures also has to be further studied to
see how the average expansion rate is related to observations.

If a realistic backreaction model does turn out to provide
acceleration in agreement with the observations, one could
say that it unifies three historically popular FRW models. There is
only matter present and the initial value of the spatial curvature
at the background level is zero as in the standard CDM model,
the universe has negative curvature as in the open CDM model,
and the expansion accelerates as in the \LCDM model.
Acceleration due to structure formation
would probably be a transient phase on the way to negative curvature
voids dominating the expansion of the universe, though the
behaviour at very late times depends on how the universe is
structured on scales which are currently far beyond the horizon,
of which we have no theoretical or observational understanding.

\ack

I thank Thomas Buchert and Ruth Durrer for discussions and
comments on the manuscript, Henk van Elst, Troels Haugb{\o}lle,
Julien Lesgourgues, David Mota, Misao Sasaki, Douglas Shaw,
Martin Sloth and Christof Wetterich for helpful discussions
and correspondence, and the Department of Physics and Astronomy
at the University of \AA{}rhus for hospitality.
This work was partly done at the CERN Theory Unit.\\

\setcounter{secnumdepth}{0}

\end{document}